\documentclass[10pt,twocolumn,pra,notitlepage,preprintnumbers,aps,nofootinbib,longbibliography,tightenlines,floatfix,superscriptaddress]{revtex4-1}

% ...
\def\Nslash{{ {\cal N}\hskip-0.55em /}}

\usepackage{hyperref}
\usepackage{mathtools}
\usepackage{amsmath,amssymb}
\usepackage{cleveref}
\usepackage{graphicx}
\usepackage{xcolor}
\usepackage{colortbl}
\definecolor{Gray}{gray}{0.9}
\usepackage[utf8]{inputenc}
\usepackage{bm}
\usepackage{placeins}
\usepackage{csquotes}
\usepackage{csquotes}
\usepackage{scalerel}
\usepackage{comment}
\usepackage[export]{adjustbox}
\usepackage{booktabs}
\allowdisplaybreaks

\newcommand{\gsim}{\raisebox{-0.7ex}{$\stackrel{\textstyle >}{\sim}$ }}

\newcount\hour \newcount\hourminute \newcount\minute
\hour=\time \divide \hour by 60
\hourminute=\hour \multiply \hourminute by 60
\minute=\time \advance \minute by -\hourminute
\newcommand{\mydate}{\ \today \ - \number\hour :\number\minute}

\makeatletter
\renewcommand*\env@matrix[1][*\c@MaxMatrixCols c]{%
  \hskip -\arraycolsep
  \let\@ifnextchar\new@ifnextchar
  \array{#1}}
\makeatother

\begin{document}

%%%%%%%%%
\title{Entanglement Spheres and a UV-IR connection in Effective Field Theories}
%%%%%%%%%
\author{Natalie Klco}
\email{natklco@caltech.edu}
\affiliation{Institute for Quantum Information and Matter (IQIM) and Walter Burke Institute for Theoretical Physics, California Institute of Technology, Pasadena CA 91125}
\author{Martin J.~Savage}
\thanks{On leave from the Institute for Nuclear Theory}
\email{mjs5@uw.edu}
\affiliation{InQubator for Quantum Simulation (IQuS), Department of Physics, University of Washington, Seattle, WA 98195}

\date{\mydate}
\preprint{IQuS@UW-21-005}

\begin{abstract}
We show that long-distance quantum correlations probe short-distance physics.
Two disjoint regions of the latticized, massless scalar field vacuum are numerically demonstrated to become separable at distances beyond the negativity sphere, which extends to infinity in the continuum limit.
The size of this quantum coherent volume is determined by the highest momentum mode supported in the identical regions, each of diameter $d$.
More generally, effective field theories (EFTs), describing a system up to a given momentum scale $\Lambda$, are expected to share this feature---entanglement between regions of the vacuum depends upon the UV-completion beyond a separation proportional to $\Lambda$.
Through calculations extended to three-dimensions, the magnitude of the negativity at which entanglement becomes sensitive to UV physics in an EFT (lattice or otherwise) is conjectured to scale as $\sim e^{-\Lambda d}$, independent of the number of spatial dimensions.
It is concluded that two-region vacuum entanglement at increasing separations depends upon the structure of the theory at increasing momentum scales.
This phenomenon may be manifest in perturbative QCD processes.
\end{abstract}

\maketitle

%%%%%%%%%%%%%%%%%
\section{Introduction}
Fundamental principles of effective field theories (EFTs) leverage clear separations of energy scales to identify relevant degrees of freedom and to build a systematically improvable
hierarchy of local operators.
By incorporating all relevant interactions consistent with the symmetries of the theory, this hierarchy accurately captures physics in regimes where scale ratios are small~(see, for example, Ref.~\cite{Weinberg:1978kz}).
While short distance properties require high energy probes (e.g., the exploration of hadronic structure through deep inelastic scattering) or quantum fluctuations (e.g.,  flavor-changing neutral currents), long distance properties may be informed by infrared (IR) observables at low energies.
Thus, long distance physics tends to be insensitive to ultraviolet (UV) modifications incorporated in an EFT through momentum truncations or the \enquote{integrating out} of high energy degrees of freedom.
In this paper, it is shown that the distillable entanglement between two disjoint regions of a massless scalar field is a long distance observable sensitive to the treatment of the UV degrees of freedom.
In particular, a finite momentum truncation, limiting the effective information resolution of the field, will cause distantly separated spatial regions of the field to not only exhibit vanishing distillable entanglement, but to become separable.
As such, momentum-space regularizations in the UV necessarily introduce an IR truncation of the vacuum quantum correlations~\cite{Reeh1961,summers1985vacuum,summers1987bell1,summers1987bell2,VALENTINI1991321,Reznik:2002fz,Reznik:2003mnx,Witten:2018zxz} at large spatial distances,  limiting the IR regime of EFT validity from the perspective of quantum mechanical inseparability.

Quantum field theory (QFT) has provided a natural unifying perspective of particles as excitations or localized packets of energy embedded in fundamental fields.
The many successes calculating entanglement in QFT benefit from an assortment of powerful strategies through lattices, replica tricks, holography, and the AdS/CFT
correspondence e.g.,~\cite{Holzhey:1994we,Callan:1994py,Audenaert:2002xfl,2004PhRvA..70e2329B,Calabrese:2004eu,Ryu:2006bv,kofler2006entanglement,Marcovitch:2008sxc,Casini:2009sr,Nishioka:2009un,Calabrese:2009ez,Calabrese:2009qy,Casini:2011kv,Calabrese:2012ew,Calabrese:2012nk,MohammadiMozaffar:2017nri,Coser_2017,Ruggiero:2018hyl,Klco:2019yrb,DiGiulio:2019cxv}.
Despite these heroic developments, particular parameter regimes
e.g., the distillable entanglement between regions distantly separated compared to their size, have evaded analytic control, retaining the importance of numerical explorations.

While the QFT description has been remarkably successful experimentally, conclusions about the underlying structure of nature are limited by the possibility that this success may result from the fact that any relativistic quantum system with Lorentz symmetry and cluster decomposition at long distances will behave as a quantum field at low energies~\cite{Weinberg:1978kz,Weinberg:1996kw}.
In fact, considerations in large volumes, inspired by entropy being non-extensive in black hole thermodynamics~\cite{Bekenstein:1973ur,Bekenstein:1974ax,Hawking:1976de,Bekenstein:1980jp,Bekenstein:1993dz}, has led to a conjecture that the QFT description egregiously overcounts degrees of freedom~\cite{tHooft:1993dmi,Susskind:1994vu}.
This perspective has inspired the proposal of a relationship between UV and IR truncations in the valid regime of an EFT---limiting the volume to avoid the extensively scaling EFT entropy from exceeding that of a potential black hole~\cite{Cohen:1998zx,Cohen:2021zzr}.
Furthermore, mixing between UV and IR physics has been connected to properties of nonlocality and noncommutative or gravitational generalizations of quantum fields, e.g.,~\cite{Minwalla:1999px,Matusis:2000jf,Minton:2007fd,Horvat:2011bs,Karczmarek:2013xxa,Lust:2017wrl,Beane:2020wjl}.

This paper demonstrates a connection between UV and IR physics with an observation that the entanglement in the vacuum of a simple quantum field, the massless noninteracting scalar field, is sensitive to high momentum modes at large spatial separations.
This extends to mixed states the speculation of Ref.~\cite{2004PhRvA..70e2329B}, informed by the modewise spatial entanglement structure of harmonic chain bipartite pure states, on the role of short wavelength modes in the persistence of vacuum entanglement.
Explicitly, we have extended numerical lattice computations of the logarithmic negativity between pixelated spherical regions to three-dimensions, allowing identification of a dimension-independent scaling of the smallest supported negativity.
Such extensions to higher dimension are non-trivial due to cancelations between polynomial- and logarithmic-scaling lattice correlation functions generating an exponentially small entanglement, manifesting as a sign problem in the lattice basis.
In the process of exploring the entanglement structure of these systems, we have provided numerical evidence, through a separability flow~\cite{PhysRevLett.87.167904}, that regions separated beyond the negativity sphere are also separable, precluding the presence of entanglement with positive partial transpose (PPT) in the massless lattice scalar field.
The array of calculations presented in this work serve as an explicit example of a more general relationship---long-distance quantum correlations probe short distance physics---broadly affecting the regime of validity for quantum observables in EFTs.

%%%%%%%%%%%%%%%%%%%%%%%%%%%%%%%%
\section{Vacuum Entanglement Sphere}
As a necessary condition for separability, the negativity~\cite{Horodecki:1996nc,Vidal:2002zz,Simon:2000zz,Plenio:2005cwa}
quantifies the violation of partial transposition---locally negating the momentum in one region for continuous variables~\cite{Simon:2000zz}---from producing a physical density matrix (non-negative eigenvalues).
If a quantum state is separable across a bipartition, partial transposition is a map that preserves the positivity of the density matrix (PPT).
Violations to this positivity herald inseparability and thus the presence of entanglement for both mixed and pure quantum states.

Consider the matrix of two-point correlation functions,
\begin{equation}
  \hat{G}_{i,j} = \langle \hat{\phi}_i \hat{\phi}_j\rangle \qquad , \qquad \hat{H}_{i,j} = \langle \hat{\pi}_i \hat{\pi}_j\rangle
  \ \ \ ,
\end{equation}
where $\{i,j\} \in A \cup B$ for two local field regions $A, B$.
For $n_s$ sites in each region, $\hat{G}$ and $\hat{H}$ are $(2n_s \times 2n_s)$-dimensional matrices with matrix elements controlled by $\mathbf{n}$, the vector separating sites $\{i,j\}$.
In the thermodynamic limit, the integral representation of the modified Bessel function of the first kind, $I_\nu(z)$, allows a succinct calculation of the necessary correlators as
\begin{equation}
  \hat{G}(\mathbf{n}) = \frac{1}{\sqrt{\pi}} \int_0^{\infty} \text{d} x \  e^{-(m^2+2D)x^2} \prod_i I_{n_i}(2x^2) \ \ \ ,
  \label{eq:phiphi}
\end{equation}
and
\begin{eqnarray}
  \hat{H}(\mathbf{n}) & = & (m^2 + 2D) G(\mathbf{n}) - \sum_{\{\mathbf{n}'\}}G(\mathbf{n}')
  \nonumber\\
  & \rightarrow &
  m^2 G(\mathbf{n}) - \nabla^2 G(\mathbf{n}) \ \ \ ,
  \label{eq:GtoHgeneral}
\end{eqnarray}
where $\{\mathbf{n}'\}$ is the set of 2D integer vectors shifted by $\pm 1$ in each direction of the $D$-dimensional lattice (see Appendix~\ref{app:correlationfunctions} for further details).
The logarithmic negativity is additive
\begin{equation}
  \mathcal{N} = -\sum_{i = 1}^{2n_s} \log_2 \min(\nu_i^\Gamma, 1) \ \ \ ,
  \label{eq:logneg}
\end{equation}
where $\nu_i^\Gamma$ are the symplectic eigenvalues of the partially transposed covariance matrix, which may be calculated as the eigenvalues of  $2\sqrt{\hat{G}\hat{H}^\Gamma}$~\cite{2004PhRvA..70e2329B,Marcovitch:2008sxc,Klco:2020rga}.
The superscript $\Gamma$ indicates partial transposition of the conjugate momentum two-point functions and may be implemented, in practice, by negating the matrix element of $\hat{H}_{i,j}$ if the sites at positions $\{i,j\}$ are in different regions $\{A,B\}$ of the field~\cite{Simon:2000zz}.

Numerical evaluations of the negativity between disjoint regions of the massless scalar field have shown that the negativity in the continuum limit with $\tilde{r} \gg d$ decays exponentially as
$\mathcal{N}\sim e^{-\beta \frac{\tilde{r}}{d}}$,
with $\tilde{r}$ the separation between the field regions and $d$ the diameter of each region.
Extractions of the negativity decay constant yield $\beta_{1D} = 2.82(3) \sim 2\sqrt{2}$~\cite{Marcovitch:2008sxc},
$\beta_{2D} = 5.29(4)$~\cite{Klco:2020rga},
and we currently estimate $\beta_{3D} = 7.6(1)$.  Appendix~\ref{app:3d} provides further discussion of this three-dimensional calculation.
As proposed in Ref.~\cite{Klco:2020rga}, this progression with dimensionality is consistent with $\beta_D\sim D$, the negativity becoming increasingly localized in higher dimensions.
At a finite lattice spacing, where regions experience finite pixelation, there exists a non-perturbative death of negativity at large separation,
$\tilde{r}_{\Nslash}$~\cite{Audenaert:2002xfl,2004PhRvA..70e2329B,kofler2006entanglement,Marcovitch:2008sxc,Calabrese:2009ez,Calabrese:2012ew,Calabrese:2012nk,MohammadiMozaffar:2017nri,Coser_2017,Klco:2019yrb,DiGiulio:2019cxv}.
The scaling of this negativity sphere with the region pixelation was previously found to be
$\tilde{r}_{\Nslash}/d \sim (\gamma/a) d$, with
$\gamma_{1D} = 1.114(2)$,
$\gamma_{2D} = 0.60(2)$~\cite{Klco:2020rga},
and we currently estimate $\gamma_{3D} = 0.43(2)$.
With a dependence of $\gamma_D \sim D^{-1}$,
a more stringent negativity sphere is observed with increasing spatial dimension.

The negativity is not generally a necessary and sufficient condition for determining the separability of Gaussian states.
In particular,
there is a form of non-distillable entanglement, bound entanglement~\cite{PhysRevLett.80.5239}, that may forbid separability while avoiding detection by the negativity criterion~\cite{1999PhRvL..82.5385B,2000PhRvA..61c0301B,PhysRevLett.85.2657,2001PhRvL..86.3658W,2001PhRvA..63c2306S,2006PhRvL..97h0501B}.
By employing the necessary and sufficient Gaussian state separability criterion of Ref.~\cite{PhysRevLett.87.167904}, which acts as a flow maintaining the separability (or non-separability) of covariance matrices while systematically simplifying the entanglement structure, it is found that regions of the scalar field outside the negativity sphere are separable. The negativity sphere is thus promoted to an entanglement sphere and describes a finite sized quantum mechanically coherent volume between regions of the field (see Appendix~\ref{app:neg} for further discussion).
As such, any observable calculated outside the entanglement sphere will be characterized by factorizable classical probability distributions.
For example, the mutual information, which persists outside this entanglement sphere, is there quantifying correlations that are entirely classical.

To the extent that spin models generically are known to exhibit vanishing two-site entanglement beyond a finite site separation~\cite{cerf2007quantum}, the presence of the entanglement sphere is not without precedence.
One intriguing implication, however, is that the long distance entanglement known to be present in the continuum quantum field is necessarily captured through the presence of genuine high-body entanglement and a Borromean structure on the lattice---regions entangled at long distances may contain vanishing negativity for all smaller subsets of sites spanning the two regions.
This reliance on genuine high-body entanglement at large distances is intuitively consistent with the complexity of low-pixelation regions being insufficient to support long-distance entanglement.

While not exact and currently limited in numerical precision due to the presence of a sign problem exacerbated in higher dimensions, the calculated dependence of the negativity decay and size of the entanglement sphere indicate that the emergence of separability at large distances is set by the UV truncation of the theory.
In particular, at the surface of the entanglement sphere, the minimum value of the negativity supported by the field before separability occurs is $\mathcal{N}_{\Nslash} \sim e^{-\beta_D \frac{\tilde{r}_{\Nslash}}{d}} \sim e^{-\beta_D \gamma_D d/a}$.
Combining the calculations above  to inform the product provides: $\beta_{1D} \gamma_{1D} = 3.14(3)$, $\beta_{2D}\gamma_{2D} = 3.2(1)$, and $\beta_{3D} \gamma_{3D} = 3.2(1)$.
The stability of this product with spatial dimension leads us to conjecture that,
\begin{equation}
  \mathcal{N}_{\Nslash} \sim e^{-\beta_D \gamma_D d/a} \sim e^{-\frac{\pi d}{a}} \sim e^{-\Lambda d} \ \ \ ,
\end{equation}
where $\Lambda$ is the scale of the UV truncation, independent of the number of spatial dimensions.
For disjoint regions of the vacuum, the threshold negativity below which the field becomes separable is determined by the diameter of each region and the highest allowed momentum mode.
In two dimensions, where $\pi d$ acts as the circumference of the circular field regions, the scaling of the negativity at the entanglement sphere is coincidentally consistent with an area dependence.
Though the negativity radius has been conventionally interpreted as a lattice artifact of no consequence to continuum physics,
this conjecture indicates that a truncation in the distillable entanglement will be present at long distances in continuum theories with finite UV truncations.

%%%%%%%%%%%%%%%%%%%%%%%%%%%%%%%
\section{Entanglement Sphere and Region Momentum}
\begin{figure}
  \includegraphics[width=0.9\columnwidth]{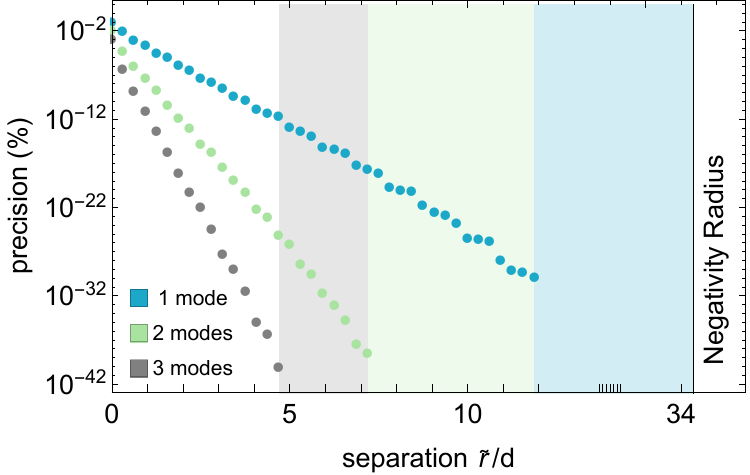}
  \caption{Precision of the mode-restricted logarithmic negativity as a function of spatial separation in the one-dimensional massless scalar field with region diameter $d = 32$.
  With increasing separation, the discrepancy vanishes and the entirety of the negativity is captured by the lowest few eigenmodes of $\hat{G}\hat{H}^{\Gamma}$, indicated by the shaded background.}
  \label{fig:negmodecontributions}
\end{figure}
\begin{figure*}
  \textbf{1D:} \begin{minipage}{0.95\textwidth}\includegraphics[width = 0.15\textwidth]{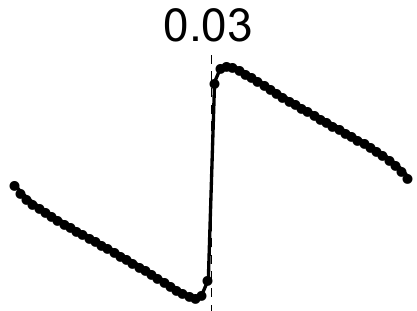}
\includegraphics[width = 0.15\textwidth]{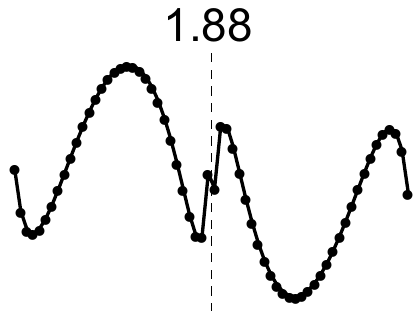}
\includegraphics[width = 0.15\textwidth]{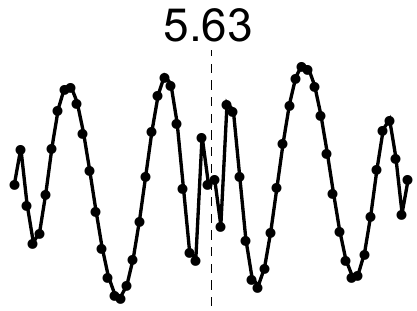}
\includegraphics[width = 0.15\textwidth]{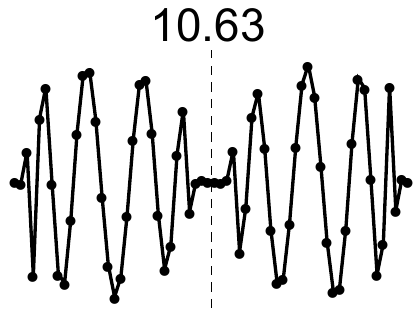}
\includegraphics[width = 0.15\textwidth]{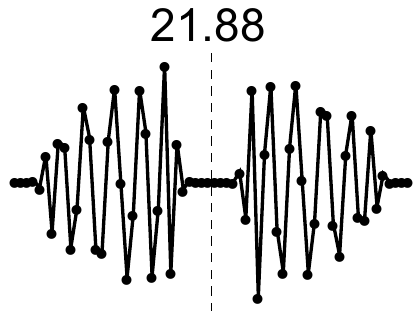}
\includegraphics[width = 0.15\textwidth]{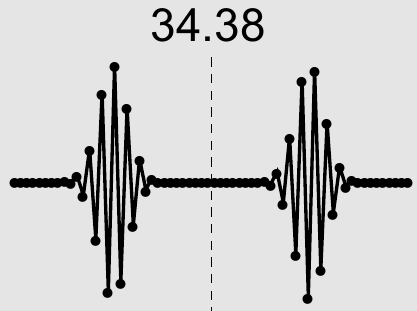}\end{minipage}\\
\textbf{2D:} \begin{minipage}{0.95\textwidth}
\includegraphics[width = 0.15\textwidth]{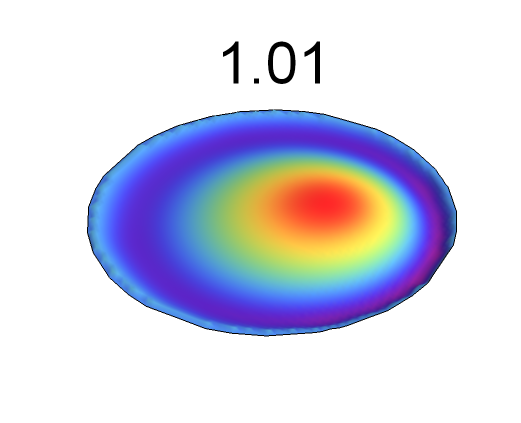}
  \includegraphics[width = 0.15\textwidth]{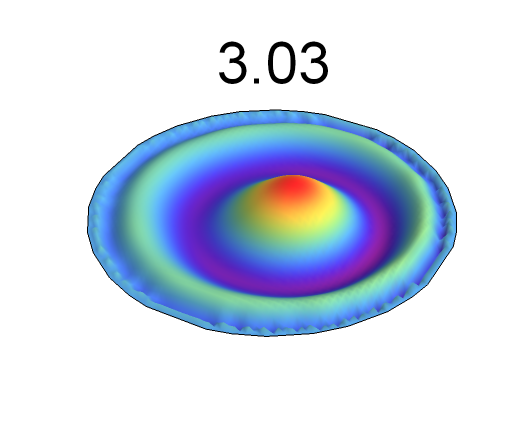}
  \includegraphics[width = 0.15\textwidth]{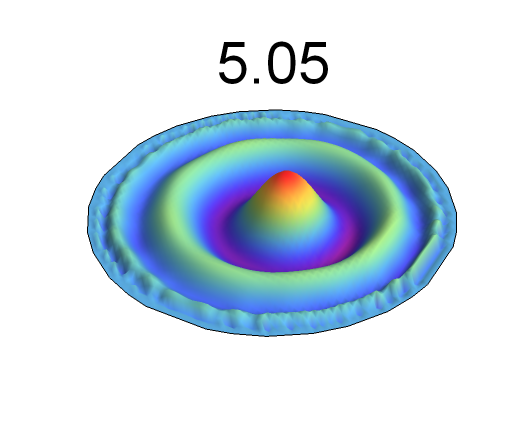}
  \includegraphics[width = 0.15\textwidth]{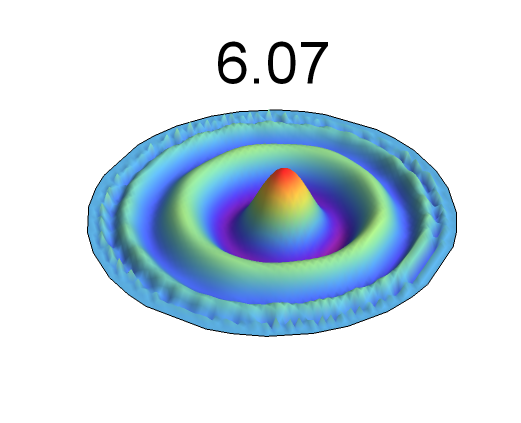}
  \includegraphics[width = 0.15\textwidth]{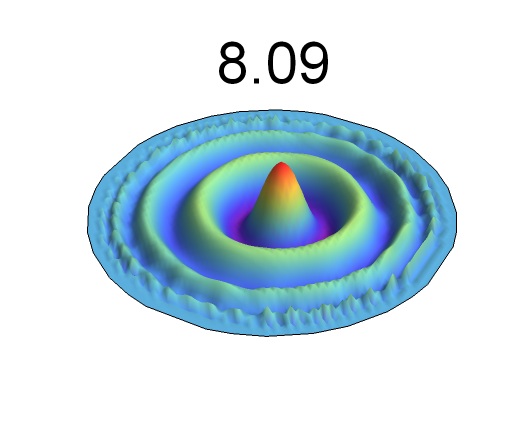}
  \includegraphics[width = 0.15\textwidth]{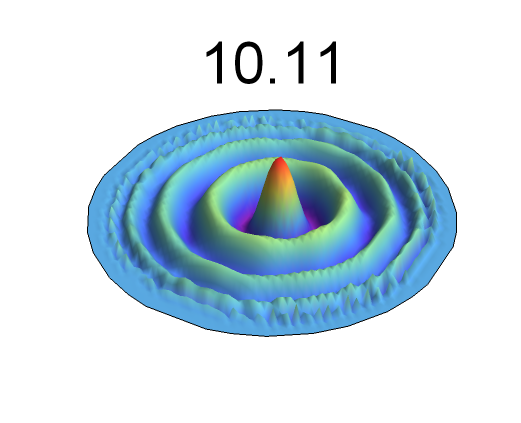}
  \end{minipage}
  \caption{Negativity ground state wavefunctions of $\hat{G}\hat{H}^\Gamma$ in the massless scalar field in one (top) and two (bottom) spatial dimensions.  The 1D wavefunctions are provided across the two regions separated at indicated distances $\tilde{r}/d$ with $d = n_s = 32$ ($\tilde{r}_{\Nslash}/d = 34.34$).  The 2D wavefunctions are shown for one region (negative spatial parity as in 1D) with  $d = 64$ for a variety of $\tilde{r}/d$ separations within the entanglement sphere, $\tilde{r}_{\Nslash}/d \sim 18$.}
  \label{fig:GHwf}
\end{figure*}
\begin{figure}
  \includegraphics[width =\columnwidth]{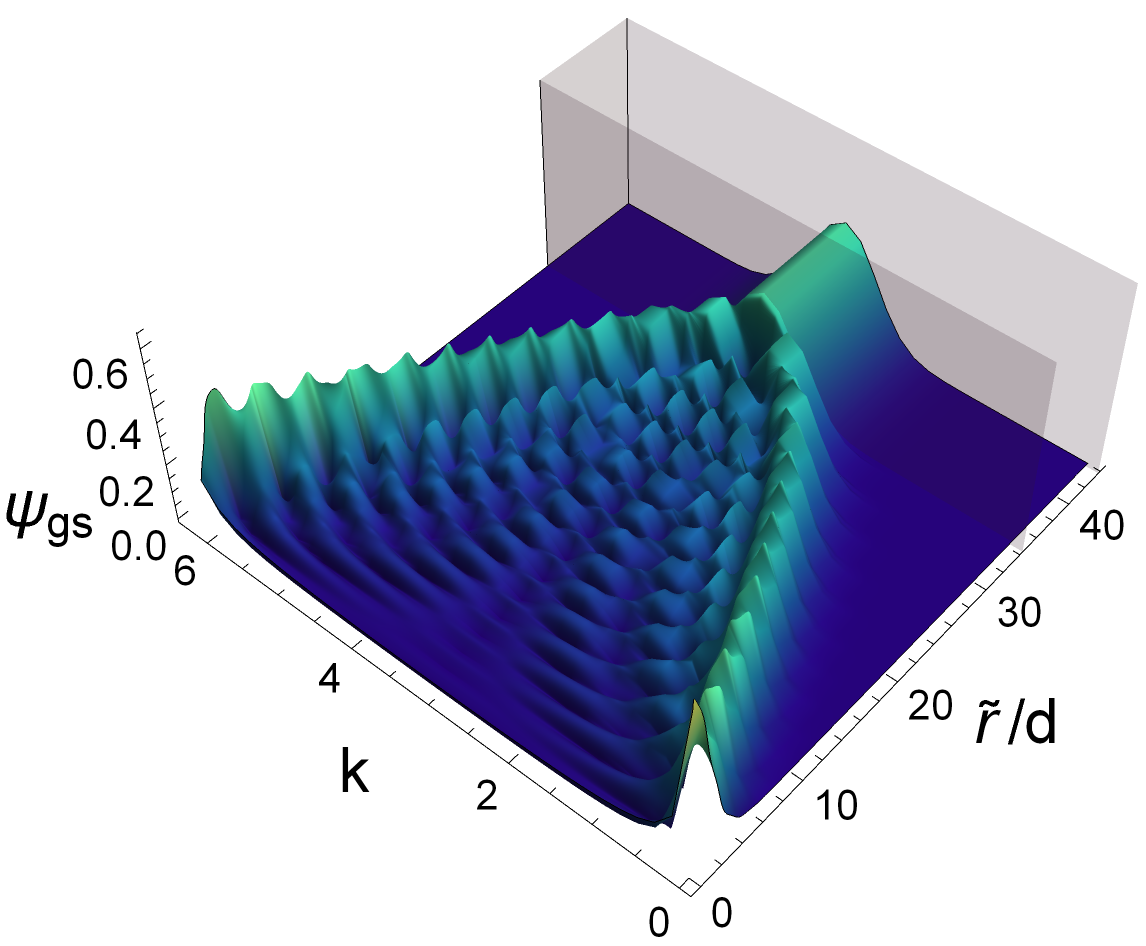}
  \caption{Momentum space, k, negativity ground state wavefunctions of $\hat{G}\hat{H}^\Gamma$ for the 1D massless scalar field isolated to a single region of $n_s = 32$ as a function of $\tilde{r}/d$ region separation.   The gray area at large separation lies outside the negativity radius, $\tilde{r}_{\Nslash}/d = 34.34$.}
  \label{fig:GHwf1D}
\end{figure}
A tangible understanding of the UV-IR connection found in the distillable entanglement at large distances can be elucidated through examination of the dominant $\hat{G}\hat{H}^\Gamma$ right-eigenvector contributing to the logarithmic negativity of Eq.~\eqref{eq:logneg}.
While many symplectic eigenvalues contribute to the negativity at small separations, this number diminishes as the separation increases, as shown in Fig.~\ref{fig:negmodecontributions}.
As the separation approaches the entanglement sphere, the negativity is characterized by the single ground state of $\hat{G}\hat{H}^\Gamma$.
In Ref.~\cite{Klco:2020rga}, we examined the negativity ground state wavefunction and found it to evolve to high momentum components with increasing separation.
The structure of the $\hat{G}\hat{H}^{\Gamma}$ ground state within the field regions at increasing separation is shown in Fig.~\ref{fig:GHwf} for one and two spatial dimensions.
In 1D, the ground state is demonstrated to have negative parity between the two regions, allowing the depiction of a single region in 2D for visual clarity.
At small separations, the negativity ground state is comprised of low-frequency contributions.
As the separation grows to multiple region diameters, fluctuations sequentially emerge from the boundary and propagate toward the central peak.
These fluctuations are comprised of high-frequency contributions and stress the pixelation of the region.
In this long-distance regime, the region negativity ground state wavefunction tends to become rotationally symmetric with a boundary condition vanishing at the edges, allowing the characterization of region entanglement to become effectively one dimensional.

The frequency space representations of the 1D region wavefunctions, $\tilde{\psi}(k) = \frac{1}{\sqrt{n_s}} \sum_{x} \psi(x) e^{i k x}$, are shown in Fig.~\ref{fig:GHwf1D} for a Brillouin zone $k \in \frac{2 \pi}{n_s} \mathbb{Z}_{n_s}$.
At small separations, support in momentum space is localized to the regimes of small region momentum at the $0$ and $2\pi$ boundaries of the Brillouin zone. As the separation between regions increases, the momentum content within the region is driven to the high momentum boundary of $k = \pi$.
Outside the entanglement sphere (gray region), the momentum-space wavefunction, now separable, locally saturates the UV truncation and ceases to evolve.

%%%%%%%%%%%%%%%%%%%%%%%%%%%%%%%%%%
\section{Implications for Effective Field Theories}
\noindent

Effective field theories provide a powerful technique for computing low-energy processes in systems with hierarchies in their energy spectrum.
The lore is that the predictions of an EFT (valid below some momentum scale $\Lambda$)
 can systematically reproduce those of the full theory (UV-complete)
when the IR degrees of freedom and symmetries match.
When the full theory is known, local counterterms in the EFT are determined by matching matrix elements computed in both theories.
The results of lattice calculations of the negativity discussed in the previous sections indicate that the behaviors of distillable entanglement and separability at long distances
of a massless non-interacting scalar field are determined by
high-momentum modes of the field, with the radius of the entanglement sphere determined by the
UV cutoff.
In particular, the exponentially-small negativity between regions of vacuum is lost beyond some
dimensionless separation determined by the maximum momentum mode allowed by the  lattice spacing, $\pi/a$.
However, this feature is generic for any low-energy EFT valid  up to some momentum scale $\Lambda$.
The higher the EFT cut off, the larger the separation between regions of the EFT vacuum that remain  entangled.

The lattice and other abrupt cutoffs in momentum space lead to a well-defined distance, beyond which
regions of the vacuum state are separable.
Other regulators can be used to render EFT computations finite  by implementing a form factor in momentum or position space e.g., Pauli-Villars (PV) or dimensional regularization (dim.\ reg.).
These provide smooth modifications of perturbative Feynman diagrams, that furnish finite loop diagrams, along with associated counterterms that depend upon the PV mass or the dim.\ reg.\ scale, $\mu$.
As calculations that can be compared with experiment are independent of the choice of regulator and renormalization scheme,
it is the UV-completion of the theory that will determine the behavior (vanishing or otherwise) of negativity at long-distances (see Appendix~\ref{app:dispersion}).
In this way, precision studies of the large-scale entanglement in the quantum vacuum can probe short-distance physics.
However, the precise translation of such entanglement studies into constraints on, for instance, beyond the standard model (BSM) physics, remains to be explored.

Rather than an IR constraint on the spatial volume scaling with the inverse UV truncation $L \sim \Lambda^{-3}$ as associated with fundamental bounds on black hole entropy~\cite{Cohen:1998zx,Cohen:2021zzr}, this work identifies a linearly scaling IR truncation, $L \sim \Lambda$, necessary to accurately capture the inseparability of the field ground state.
While the core principles of these cutoffs lie in the saturation of spatially localized information, their distinct scaling suggests a stringency crossover---the formation of black holes
being the relevant constraint for UV truncations above the crossover and the separability criterion being the relevant constraint for UV truncations below.
To gain some insight into the potential impacts, we provide mass- and length-scales for two different scenarios.
Using, in three dimensions, $\tilde{r}_\Nslash \sim d^2/(3 a)\rightarrow \Lambda d^2/(3 \pi)$, and setting
$\Lambda=M_{\rm pl}$ to be the Planck mass,
we find that regions of vacuum of
a massless noninteracting scalar field
approximately the size of the proton
are separable at distances beyond $\sim 6~{\rm km}$,
and that regions approximately the size of an atom are separable beyond $\sim 10^{11}~{\rm km}$.
If the cutoff of the EFT is instead $\Lambda\sim 1~{\rm TeV}$, then
proton sized regions are separable beyond $\sim 500~{\rm fm}$ and atomic sized regions beyond
$\sim 5~{\rm mm}$.
Evaluating the potential for possible signatures from these distance scales in experiment, one takes pause from
the values of negativity at the point of separability, which are
$\sim 10^{-1800}$ and $\sim 10^{-2 \times 10^8}$, respectively, for a TeV scale cut off.
The magnitude of this effect falls exponentially with separation, imposing what are likely to be severe limits
on the constraints that can be determined by experiment.

%%%%%%%%%%%%%%%%%%%%%%%%%%%%%%%%%%
\section{Discussion}
\noindent
By considering massless, non-interacting scalar field theory,
the  onset of separability between regions of a quantum vacuum in discretized systems has been shown to
result from a connection between long-distance and short-distance physics.
In particular, the maximum distance between regions of the vacuum that are entangled is directly related
to the high-momentum modes of the field.
For a lattice field theory, this distance is set by the inverse lattice spacing.
While analytic results and a deep understanding of the underlying mechanism(s) remain to be uncovered, numerical explorations indicate that
the onset of separability is connected to a saturation of information contained in regions of the pixelated vacuum.
This naturally suggests that such effects are present in the vacuum of any quantum field theory that
has a UV cutoff, as in the case of any EFT.
This violates EFT lore, which assumes that IR physics in a complete theory can be systematically recovered in a low-energy EFT.
The present results show, however, that a UV cutoff in an EFT leads to separable regions of the vacuum beyond an entanglement sphere.
Further lattice calculations are required to reduce uncertainties associated with the negativity parameters, and to pursue calculations in higher spatial dimensions to better define their dimensional scaling.

For most systems, the discrepancies in measures of entanglement are very small, with estimates suggesting they will be challenging to explore experimentally.
However, recent advances in the control of quantum systems capable of representing bosonic fields e.g.,~\cite{Wang:2018kla,Holland:2019zju,Roy:2020ppa,hung2021quantum} and the potential to leverage quantum computing technologies as quantum detectors of background field entanglement properties yields numerous directions for potential experimental connection.
Beyond the vacuum, experimental signatures of this UV-IR connection, and its implications for low-energy EFTs, may be more profitably sought in excited states of the field theory, in particular, in systems of two or more wavepackets, themselves amenable to detection and measurement.
For example, in the EFT description of S-channel scattering of nuclei, entanglement power has been proposed to impact the relevant hierarchy of local operators~\cite{Beane:2018oxh,Beane:2020wjl}.

While fundamental massless scalar fields do not exist in nature, fields with massless or light excitations and modest or weak interaction strengths e.g., electromagnetism, perturbative QCD, nuclear EFTs around the chiral limit~\cite{Fleming:1999ee,Beane:2001bc},  gravity, axions, Bose-Einstein condensates~(e.g., Refs.~\cite{2018Sci...360..413K,Sanchez-Kuntz:2020lux}), and neutrinos, may exhibit a modified vacuum entanglement structure at long distances due to UV physics.
High energy processes in the nucleon or nucleus, probing distances below the confinement scale with a spatial momentum transfer ${\bf Q}$ and UV cutoff $\Lambda$, may be sensitive~\cite{Kharzeev:2017qzs,Baker:2017wtt,Berges:2018cny,Tu:2019ouv,Kharzeev:2021yyf} to such modifications for $\tilde{r}_{\Nslash} \ \gsim \Lambda/(3\pi |{\bf Q}|^2)$, resulting in a negativity deviation $\sim \exp(-\Lambda/|{\bf Q}|)$.
Similar signatures are also expected in massive fields, where the long-distance negativity decay becomes Gaussian~\cite{Marcovitch:2008sxc}.
It would appear that the phenomenology of entanglement in the quantum vacuum at long distances
may depend on what lies beyond the standard model describing electroweak and strong interactions.

%%%%%%%%%%%%%%%%%%
\begin{acknowledgments}
We would like to thank Silas Beane, Douglas Beck, Roland Farrell, David Kaplan, Aidan Murran, John Preskill, and Alessandro Roggero for valuable discussions.
We have made extensive use of Wolfram Mathematica~\cite{Mathematica} and the Avanpix multiprecision computing toolbox~\cite{mct2015} for MATLAB~\cite{MATLAB:2020}.  Numerical results are available upon request. NK is supported in part by the Walter Burke Institute for Theoretical Physics, and by the U.S.
Department of Energy Office of Science, Office of Advanced Scientific Computing Research, (DE-SC0020290), and
Office of High Energy Physics DE-ACO2-07CH11359. MJS was supported in part by the U.S. Department of Energy,
Office of Science, Office of Nuclear Physics, InQubator for Quantum Simulation (IQuS) under Award Number DOE
(NP) Award DE-SC0020970.
\end{acknowledgments}

\bibliography{sclrscpbib}

%%%%%%%%%%%%%%%%%%
\onecolumngrid
\appendix

%%%%%%%%%%%%%%%%%%
\section{Lattice Correlation Functions in the Thermodynamic Limit}
\label{app:correlationfunctions}

For free scalar fields, two-point correlation functions  define the information content in the field.
In this appendix,
formulations of these expectation values are presented in the thermodynamic limit of large lattices with a form conducive to computation.
In 1D and 2D, analytic expressions are provided, while in higher dimensions,
a one dimensional integral is provided with  convergence that improves with increasing dimension.
For  $D\geq 2$, correlation functions are IR convergent, allowing the mass to be reliably set to zero.

For arbitrary $D$,
the two-point lattice field correlation function with unit lattice spacing and field separation described by  a vector of positive integers,
$\mathbf{n}$, is
\begin{equation}
  \langle \phi(0) \phi(\mathbf{n}) \rangle = G(\mathbf{n}) = \sum_{p} \frac{ e^{i \mathbf{p}\cdot \mathbf{n}} }{2 L^D} \frac{1}{ \sqrt{m^2 + 4 \sum_{i} \sin^2\left( \frac{p_i}{2} \right) } }
  \ \ ,
\end{equation}
for $p_i \in \frac{2\pi}{L} \mathbb{Z}_L$ with $\mathbb{Z}_L$ the set of integers between $0$ and $(L-1)$.
In the thermodynamic limit ($L \rightarrow \infty$),
\begin{equation}
  G(\mathbf{n}) =
  \frac{1}{{2\pi^D }}\int_0^\pi \text{d}^D  p \    \frac{\prod_i \cos(p_i n_i)}{ \sqrt{m^2 + 4 \sum_{i} \sin^2\left( \frac{p_i}{2} \right) } }
  \ \ \ ,
\end{equation}
with $\text{d}^d p = \prod_i \text{d} p_i$.
A Gaussian integral may be introduced to capture the radical in the denominator
\begin{equation}
  G(\mathbf{n}) = \frac{1}{{2\pi^D }} \int_0^\pi \text{d}^D  p \ \prod_i \cos(p_i n_i)
  \left[ \frac{2}{\sqrt{\pi}} \int_0^{\infty} \text{d} x \ e^{-(m^2+ 2 D)x^2}e^{\sum_{i}(2\cos p_i) x^2}  \right]
  \ \ \ .
\end{equation}
The integral representation of the modified Bessel function of the first kind may subsequently be identified to yield,
\begin{equation}
  G(\mathbf{n}) = \frac{1}{\sqrt{\pi}} \int_0^{\infty} \text{d} x \  e^{-(m^2+2 D )x^2} \prod_i I_{n_i}(2x^2)
  \ \ \ ,
\end{equation}
where, for integer order, the modified Bessel function may be expressed as,
\begin{equation}
  I_{\alpha}(z) = \frac{1}{\pi} \int_0^\pi \text{d} \theta \  e^{z \cos \theta} \cos \alpha \theta
  \ \ \ .
  \label{eq:BesselIintegral}
\end{equation}
The resulting expression for the two-point correlation function is a single integral for any $D$.
As $I_\alpha(z) \sim e^z/\sqrt{2\pi z}$ for large arguments, the convergence of these integrals improves with increasing $D$.

Employing similar procedures for the two-point functions of conjugate momentum field operators
yields analogous computationally advantageous expressions.
In the thermodynamic limit with massless fields, the expectation value may be written as,
\begin{align}
  \langle \pi(0) \pi(\mathbf{n}) \rangle &\equiv H(\mathbf{n}) \\
  &= \frac{1}{{2\pi^D }}\int_0^\pi \text{d}^D  p \
  \prod_i \cos(p_i n_i)  \sqrt{m^2 + 4 \sum_{i} \sin^2\left( \frac{p_i}{2} \right) }  \\
  &= \frac{1 }{{2\pi^D }}\int_0^\pi \text{d}^D  p \
  \frac{\prod_i \cos(p_i n_i)}{ \sqrt{m^2 + 4 \sum_{i} \sin^2\left( \frac{p_i}{2} \right) } } \left(m^2 + 2 D  - 2\sum_j \cos(p_j) \right)
  \ \ \ .
\end{align}
Associating the additional $\cos(p_i)$ factors with a partial derivative of the modified Bessel function, Eq.~\eqref{eq:BesselIintegral}, with respect to its argument, the correlator becomes
\begin{equation}
  H(\mathbf{n}) = \int_0^{\infty} \frac{e^{-(m^2 + 2D)x^2}}{\sqrt{\pi}} \left( (m^2+2D) \prod_i I_{n_i}(2x^2) - 2 \sum_j \frac{\partial_j}{\partial (2x^2)} \prod_i I_{n_i}(2x^2) \right) \ \ \ ,
\end{equation}
where $\frac{\partial_j}{\partial (2x^2)} $ indicates the argument derivative of the Bessel function associated with the $j^{\text{th}}$ lattice direction.
Given that this  is expressible as the average of the Bessel functions of neighboring order,
\begin{equation}
  \partial_z I_\alpha(z) = \frac{I_{\alpha-1}(z) + I_{\alpha+1}(z)}{2} \ \ \ ,
\end{equation}
$H(\mathbf{n})$ can be written as a linear combination of field correlators,
\begin{eqnarray}
  H(\mathbf{n}) & = & (m^2 + 2D) G(\mathbf{n}) - \sum_{\{\mathbf{n}'\}}G(\mathbf{n}')
  \ \ \ ,
  \nonumber\\
  & \rightarrow &
  m^2 G(\mathbf{n}) - \nabla^2 G(\mathbf{n})
  \  \ ,
\end{eqnarray}
where the set $\{\mathbf{n}'\}$ is the set of 2D integer vectors shifted by $\pm 1$ in each possible direction of the $D$ dimensional lattice.
Thus, the numerical stability of the field correlators  extends to correlators in conjugate momentum space.

%%%%%%%%%%%%%%%%%%%%%%%%%%%%%%%%%%%%%%%%%%%%%
\subsection{One Spatial Dimension}
The integral expression of Eq.~\eqref{eq:phiphi} in 1D can be evaluated in closed form as,
\begin{equation}
  G(n) = \frac{1}{2 \sqrt{\pi} (2 + m^2)^{n+1/2}} \Gamma\left[ \frac{1}{2} + n\right] {}_2\tilde{F}_1 \left( \begin{matrix}
    \frac{1+2n}{4}, \frac{3+2n}{4} \\
    1+n
  \end{matrix}
  ; \frac{4}{(2+m^2)^2} \right)
  \ \ ,
\end{equation}
where
${}_2\tilde{F}_1\left( \begin{matrix}
  a, b\\c
\end{matrix} ; z\right) ={}_2F_1\left( \begin{matrix}
  a, b\\c
\end{matrix} ; z\right) /\Gamma(c)$
is the regularized hypergeometric function, consistent with e.g.,~Ref.~\cite{Coser_2017}.  The ${}_2F_1$ function is finite for finite values of its parametric arguments and for $|z|<1$.
Thus, the mass in 1D is a necessary IR regulator that must be maintained nonzero throughout  calculations.

%%%%%%%%%%%%%%%%%%%%%%%%%%%%%%%%%%%%%%%%%%%%%
\subsection{Two Spatial Dimensions}
In 2D, the field correlators can  be solved analytically, a valuable capability for the finely-pixelated wavefunctions
presented in Fig.~\ref{fig:GHwf} of the main text.
Using the following identity for the product of two modified Bessel functions of equal
argument~\cite[Eq.~10.31.3]{NIST:DLMF},
the two point function of Eq.~\eqref{eq:phiphi} becomes,
\begin{equation}
  G(n_x, n_y) = \frac{1}{\sqrt{\pi}} \int_0^\infty \text{d} x \ e^{-(m^2+4)x^2}   x^{2(n_x + n_y)} \sum_{k = 0}^{\infty} \frac{\left( n_x + n_y + k + 1\right)_k x^{4k}}{k! \Gamma(n_x + k + 1) \Gamma ( n_y + k + 1)}
  \ \ \ ,
\end{equation}
relying on the closure of a product of ${}_0F_1$ functions into a ${}_2F_3$ generalized hypergeometric function.
The $x$ integral is  calculated as a Gaussian moment,
\begin{equation}
  G(n_x, n_y) = \frac{1}{2 (4+m^2)^{\frac{1+2n_x+2n_y}{2} } \sqrt{\pi}}  \sum_{k = 0}^{\infty} \frac{\left( n_x + n_y + k + 1\right)_k}{k! \Gamma(n_x + k + 1) \Gamma ( n_y + k + 1)} \frac{\Gamma\left( \frac{1}{2} + 2k + n_x + n_y\right)}{(4+m^2)^{2k} } \ \ \ .
\end{equation}
Using the Legendre duplication relation of the Gamma function and the associated Pochhammer identity,
\begin{eqnarray}
  \left( \frac{z}{2} \right)_n \left( \frac{z}{2} + \frac{1}{2} \right)_n
  & = & 2^{-2n} (z)_{2n}
 \ \ \ ,\ \ \
     \frac{\left( \frac{a+b+1}{2}\right)_k \left( \frac{a + b+ 2}{2} \right)_k}{ \left(a+b+1\right)_k}
    \ =\  4^{-k} (1+a+b+k)_k
    \ \ \ ,
  \label{eq:Pid}
\end{eqnarray}
a generalized hypergeometric function can be identified for the two-point correlation function,
\begin{multline}
  G(n_x, n_y) = \frac{1}{2  \sqrt{\pi}(4+m^2)^{1/2+n_x+n_y}} \frac{\Gamma\left( \frac{1}{2} + n_x + n_y\right) }{\Gamma(n_x+1) \Gamma(n_y+1)}  \\ \times {}_4F_3\left( \begin{matrix}
    \frac{n_x+n_y+1}{2}, \frac{n_x+n_y+2}{2}, \frac{n_x+n_y}{2} + \frac{1}{4}, \frac{n_x+n_y}{2} + \frac{3}{4} \\
    1+n_x, 1+n_y, 1+n_x+n_y
  \end{matrix} ; \frac{16}{(4+m^2)^2} \right)
  \ \ \ ,
\end{multline}
where
\begin{equation}
  {}_pF_q\left( \begin{matrix}
    a_1, \cdots, a_p \\
    b_1, \cdots, b_q
  \end{matrix} ; z \right) \equiv \sum_{k = 0}^{\infty} \frac{z^k}{k!} \frac{\prod_{j = 1}^p (a_j)_k}{\prod_{\ell = 1}^q (b_\ell)_k }
  \ \ \ .
\end{equation}
In the massless limit, the argument of the hypergeometric function becomes unity.
Because the $|z|=1$ case is finite and well defined for
${}_4F_3\left( \begin{matrix}
  a_1, a_2, a_3, a_4 \\ b_1, b_2, b_3\end{matrix};z\right) $
 if $\sum_j b_j - \sum_j a_j > 0$ (which will always be satisfied above),
 the mass does not need to be retained as an explicit IR regulator.
The correlators of conjugate momentum operators are related through Eq.~\eqref{eq:GtoHgeneral}  as,
\begin{multline}
  H(n_x,n_y) = (4+m^2) G(n_x, n_y) - G(n_x-1, n_y) - G(n_x+1, n_y) - G(n_x,n_y-1) - G(n_x, n_y+1)
  \ \ \ .
\end{multline}

It is expected that analytic expressions exist for the field correlators in higher dimension,
$D > 2$, though their form remains elusive.
For 3D, it has been computationally practical (leveraging the mentioned improving convergence in higher dimensions) to calculate the lattice two-point functions numerically through a truncated expansion  of  modified Bessel functions in Eq.~\eqref{eq:phiphi}.

%%%%%%%%%%%%%%%%%%%%%%%%%%%%%%%%%%%%%%%%%%%%%
\section{Logarithmic Negativity, Gaussian Separability Criteria and the Entanglement Sphere}
\label{app:neg}

While logarithmic negativity can be difficult to interpret as an entanglement measure in general, its current application, combined with the separability criterion of Ref.~\cite{PhysRevLett.87.167904} and the distillation criterion of Ref.~\cite{giedke2001distillability}, has yielded a uniquely informative quantification of quantum correlations in the free lattice scalar field theory vacuum.
Generically,  negativity has been connected to the operational measure of \emph{distillable entanglement}
i.e., the population of entangled pairs that can be asymptotically extracted from an ensemble of the state through  local interactions~\cite{Vidal:2002zz}.
Thus, the negativity will inform the spacelike entanglement that could be extracted by locally interacting detectors.

As an upper bound to the distillable entanglement, a vanishing logarithmic negativity between regions of a
field indicates that two sensors/detectors interacting at spacelike separations with the field would never themselves become entangled,
even if the local regions of the field are expressed with a single non-separable quantum wavefunction.
The latter extension indicates that a vanishing negativity is generically inconclusive with respect to the separability
of the regions (e.g., due to the possibility of bound entanglement~\cite{PhysRevLett.80.5239}).
Non-zero negativity, however, is conclusive with respect to inseparability.
Because negativity quantifies the physical validity of a density matrix after transformation by a particular local positive map (partial transposition), any violation of physicality heralds the presence of entanglement.
However, as an upper bound, a non-zero negativity generically carries no conclusive information about the operational entanglement structure of the field.

For the specific application to Gaussian states representing the free scalar vacuum,
both vanishing and non-vanishing negativity calculations become more informative.
As such, the free scalar vacuum is a special case, distinct but reminiscent of e.g., the $(1\times N)$-mode or bisymmetric Gaussian states for which the negativity is known to be a necessary and sufficient condition for entanglement.
For Gaussian states, if a bipartite system is found to have non-zero negativity (NPT), then non-zero distillable entanglement is determined to be present~\cite{giedke2001distillability}.

\begin{figure}
\textbf{flow step:} \hspace{0.89\textwidth} \
\\
\textbf{0} \hspace{0.095\textwidth} \textbf{1} \hspace{0.095\textwidth} \textbf{2} \hspace{0.095\textwidth} \textbf{3} \hspace{0.095\textwidth} \textbf{4} \hspace{0.095\textwidth} \textbf{5} \hspace{0.095\textwidth} \textbf{6} \hspace{0.095\textwidth} \textbf{7} \qquad \\
\includegraphics[width = 0.11\textwidth]{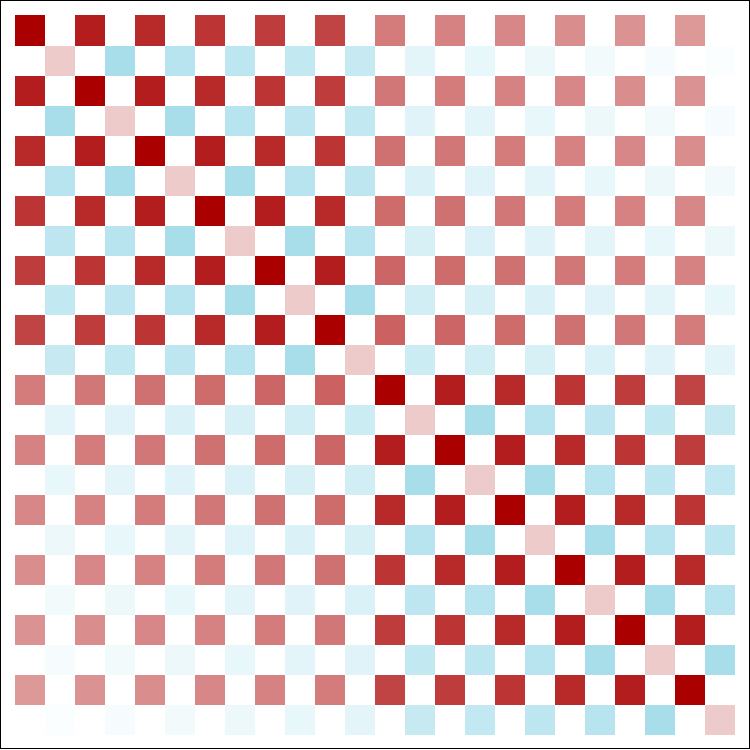}
\includegraphics[width = 0.11\textwidth]{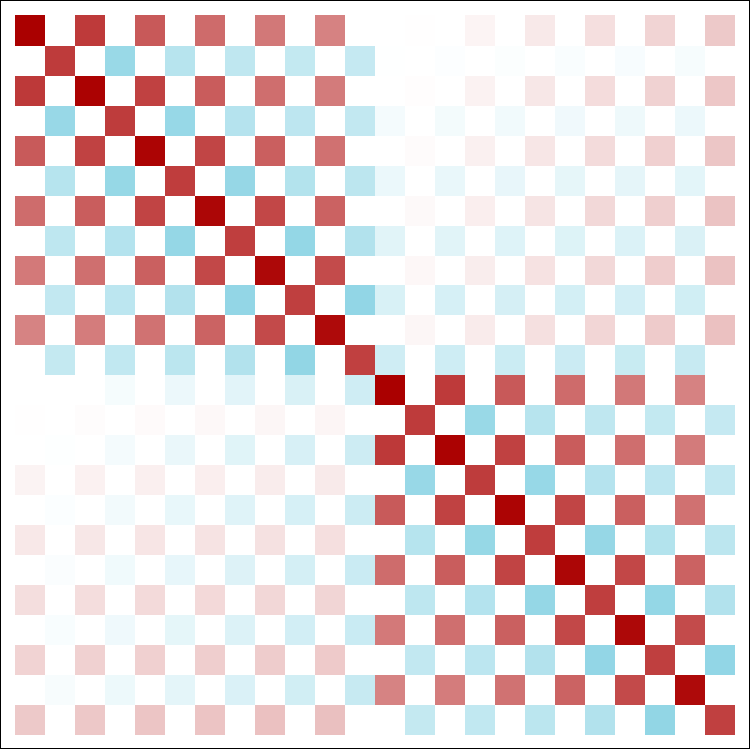}
\includegraphics[width = 0.11\textwidth]{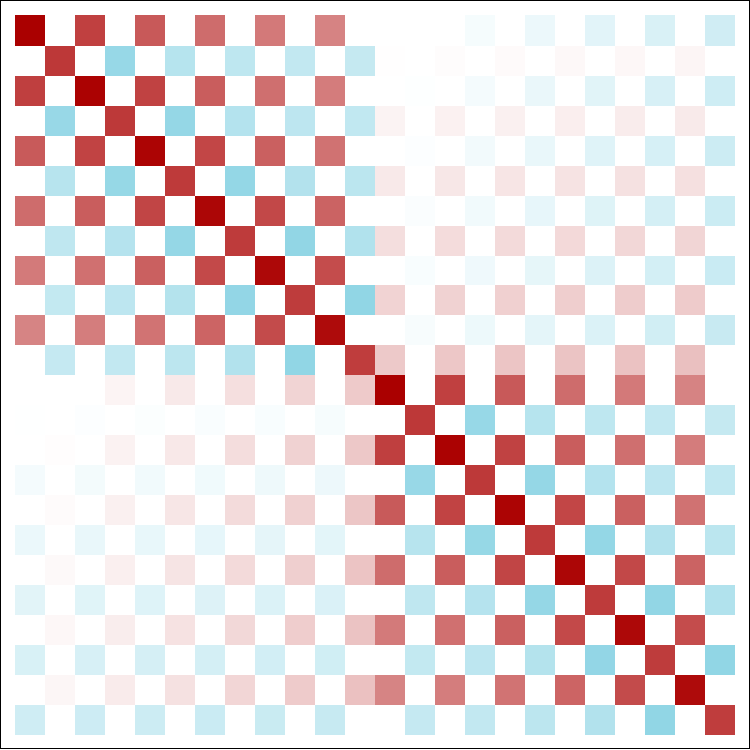}
\includegraphics[width = 0.11\textwidth]{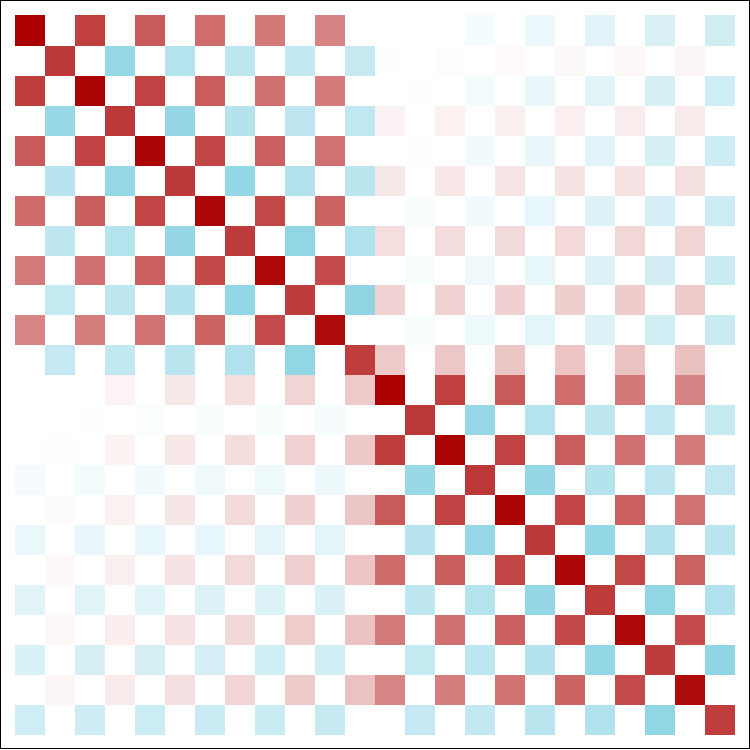}
\includegraphics[width = 0.11\textwidth]{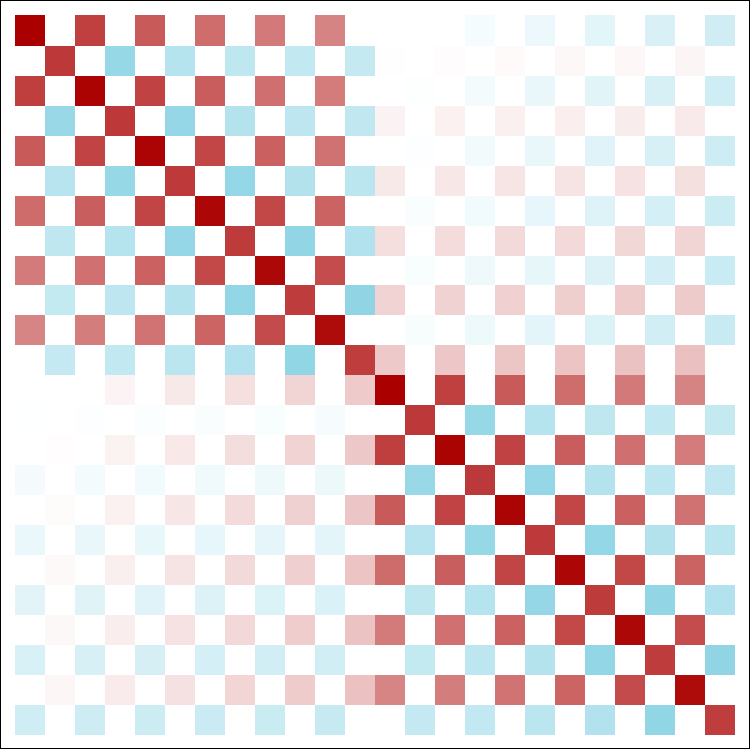}
\includegraphics[width = 0.11\textwidth]{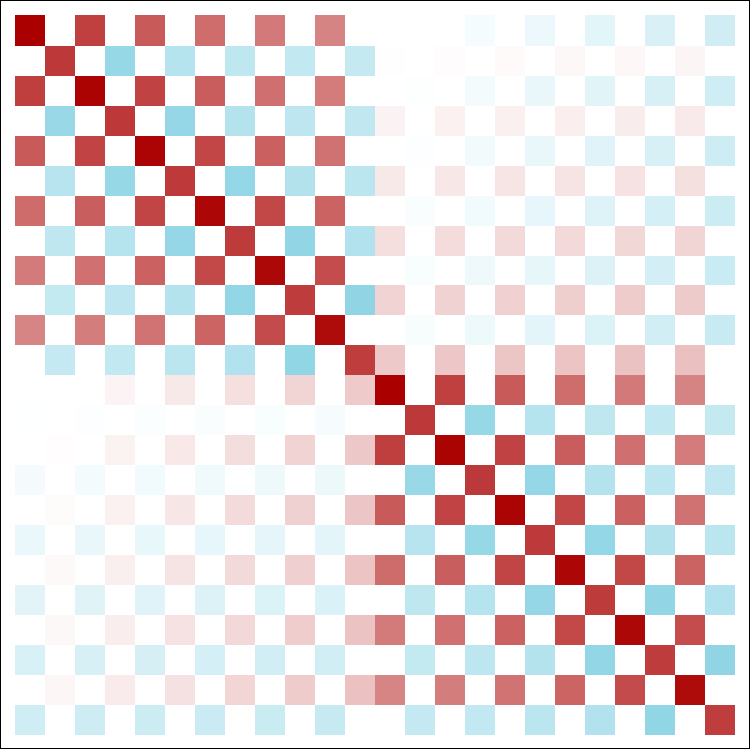}
\includegraphics[width = 0.11\textwidth]{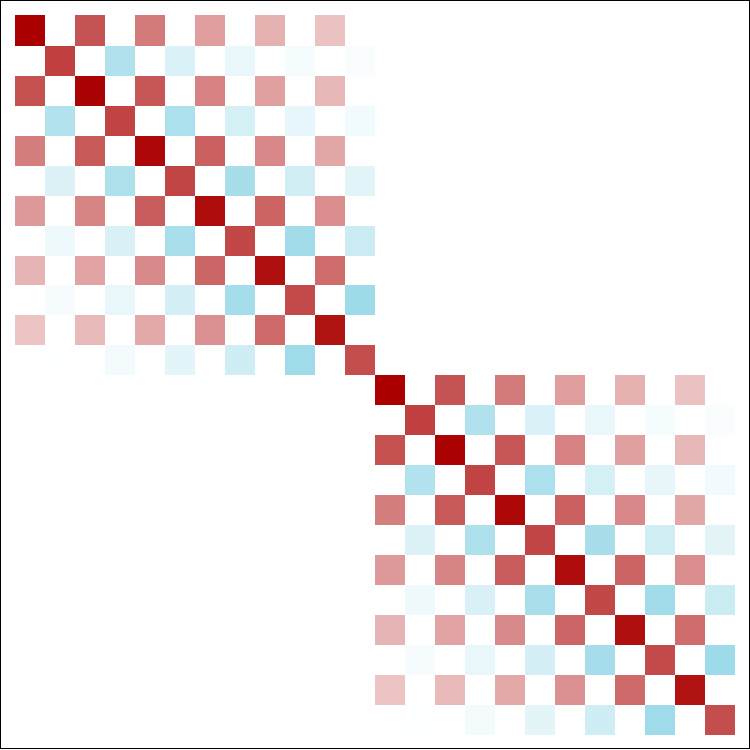}
\includegraphics[width = 0.11\textwidth]{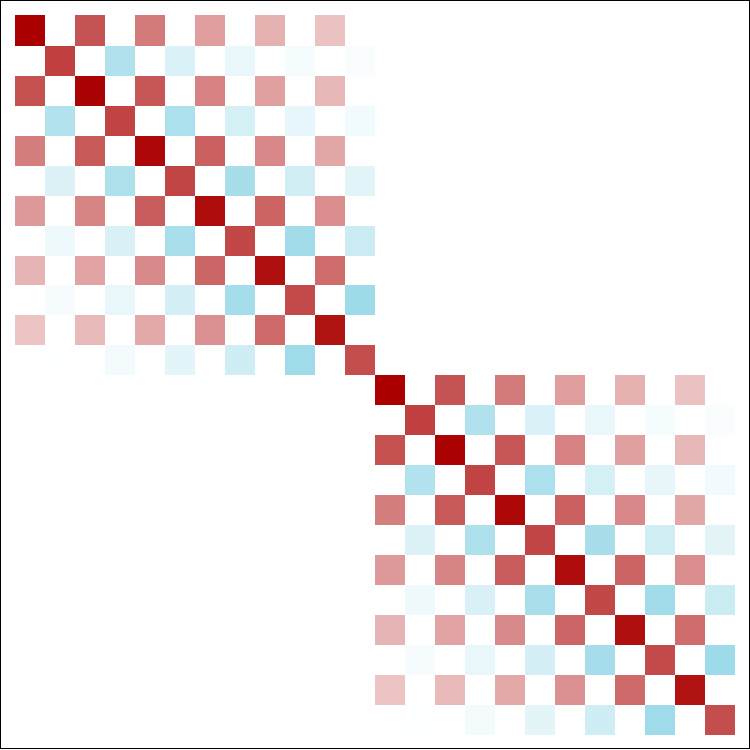}
\textbf{conclusion:} \hspace{0.87\textwidth} \ \\
\textbf{?} \hspace{0.092\textwidth} \textbf{?} \hspace{0.092\textwidth} \textbf{?} \hspace{0.092\textwidth} \textbf{?} \hspace{0.092\textwidth} \textbf{?} \hspace{0.076\textwidth} \textbf{sep} \hspace{0.075\textwidth} \textbf{sep} \hspace{0.075\textwidth} \textbf{sep} \qquad \\
\rule{8cm}{0.4pt}
\\
\includegraphics[width=0.66\textwidth]{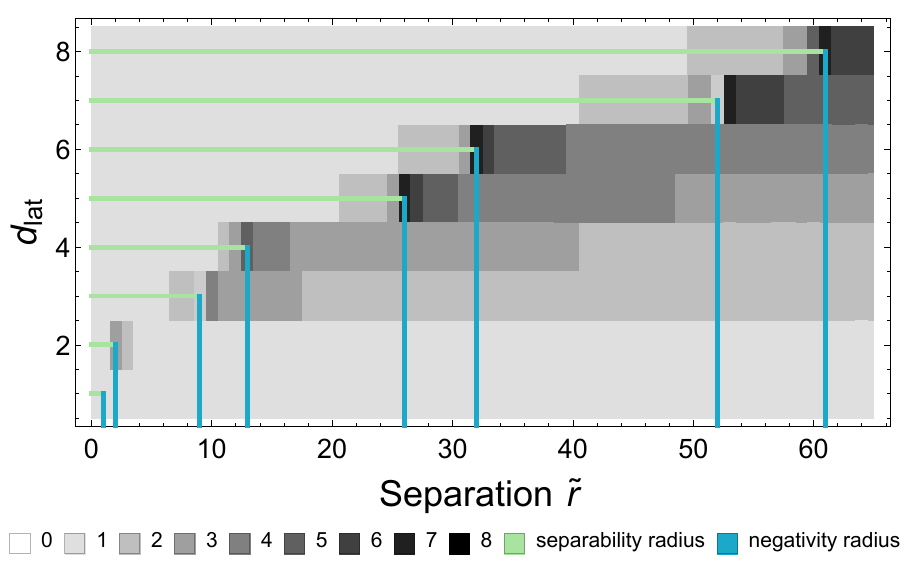}
\caption{
(Upper panel)
Steps in the covariance matrix flow arising from two regions of 1D lattice scalar field theory vacuum with $d_{\rm lat} = 6$
for a representative separation with $\tilde{r} = 34$ outside the separability sphere, as defined in Eq.~(\ref{eq:covflow}).
(Lower panel)
The separability and negativity radii are shown to coincide for a range of $d_{\rm lat}$.
}
\label{fig:separabilityflow}
\end{figure}
To further analyze the systems beyond the negativity sphere with vanishing negativity (positive partial transpose, PPT), the entanglement criterion of Ref.~\cite{PhysRevLett.87.167904} can be profitably employed.
A \enquote{flow} for the vacuum state covariance matrix, $\sigma$, was designed of the following form
\begin{equation}
  \sigma_0 = \begin{pmatrix}
    A_0 & C_0 \\
    C_0^T & B_0
  \end{pmatrix}
  \ \ \ ,
  \qquad \sigma_{N+1} =
  \begin{pmatrix}
    A_N-\text{Re}(X_N) & -\text{Im}\left(X_N\right) \\
    -\text{Im}\left(X_N^T\right) &  A_N-\text{Re}(X_N)
  \end{pmatrix}
  \ \ \ ,
  \label{eq:covflow}
\end{equation}
with $X_N = C_N(B_N - i \Omega)^{-1} C_N^T$ inspired by the Schur complement's relation to the matrix positivity condition.
For the $N$ continuous variable modes of the field,
$i\Omega=[\hat{\mathbf{r}}, \hat{\mathbf{r}}^T]$ is the matrix of canonical commutation relations incorporating the
uncertainty principle into the symplectic language with $\hat{\mathbf{r}}$ the $2N$ dimensional vector of $\hat{\phi}_i, \hat{\pi}_i$
canonical operators.
It is shown in Ref.~\cite{PhysRevLett.87.167904} that this non-linear map preserves the separability of states, flowing (in)separable states to (in)separable states.
In addition to this preservation, the flow is demonstrated to transform covariance matrices into forms for which the
separability criterion can be more readily determined by calculating e.g., the physicality of the reduced covariance
matrix of the first region, $A_N$, indicating inseparability if $A_N$ fails to pass the uncertainty principle
requirements of $\sigma_{\rm phys} - i \Omega \geq 0$.
While the two criteria used to \enquote{detect} separability or inseparability of $\sigma$ are commonly
inconclusive when applied to covariance matrices, a few steps of the separability-preserving flow in
practice produce a $\sigma_N$ whose separability condition can be conclusively determined,
and thus that of $\sigma_0$ determined by extension.

For example, consider the case of 6-site field regions $(d_{\rm lat} = 6)$ in 1D of the massless non-interacting
lattice scalar field.
As previously calculated (e.g.,~Ref.~\cite{Klco:2020rga}) the first lattice separation at which the negativity vanishes for this system, defining the negativity radius, is $\tilde{r}_{\Nslash} = 32$.
Inside this radius, the non-zero negativity indicates inseparability, but at and outside this separation, the vanishing negativity does not inform the separability of the field regions.
Employing the separability-preserving non-linear map~\cite{PhysRevLett.87.167904}
in Eq.~(\ref{eq:covflow})
for a representative separation outside the negativity sphere of $\tilde{r} = 34$
produces the covariance matrix flow shown at the top of Fig.~\ref{fig:separabilityflow}.
The initial covariance matrix at flow step 0, written in the basis $\hat{\mathbf{r}} = \{\hat{\phi}_0, \hat{\pi}_0, \hat{\phi}_1, \hat{\pi}_1, \cdots, \hat{\phi}_5, \hat{\pi}_5\}$,
is inconclusive
with regard to the separability and inseparability criteria.
However, after 5 flow steps, the covariance matrix $\sigma_5$ is determined to be separable.
In this case, the flow arrives at the clearly separable fixed point in which the off-diagonal covariance matrix elements
between regions vanish, though the criterion is capable of identifying separability prior to this point.
The separability of $\sigma_5$ implies (by design) also the separability of $\sigma_0$,
indicating that the vanishing negativity in this example is the result of a stronger condition that the two regions of the field are separable.

Exploring further this flow and the relationship between negativity and separability in the lattice scalar field ground state,
the lower panel of Fig.~\ref{fig:separabilityflow} shows a gray-scale distribution of the number of flow steps
necessary to conclusively determine separability.
It can be clearly seen that the number of flow steps necessary to determine separability/inseparability is maximized along the surface of the negativity sphere.
Unfortunately, the correlation matrices explored throughout the flow evolution become poorly conditioned at increased pixelations, requiring careful control of numerical precision even far from the continuum.
Furthermore, Fig.~\ref{fig:separabilityflow} shows
that the negativity radii (vertical blue lines), as calculated from Ref.~\cite{Klco:2020rga} for each $d_{\rm lat}$ indicated by the line's height, exactly agree with the separability radii (length of horizontal green lines), determined by flowing each covariance matrix of region diameter $d_{\rm lat}$ and lattice region separation $\tilde{r}$.
As discussed in the main text, this comparison indicates that the free lattice scalar field theory
vacuum is a non-generic quantum state for which vanishing negativity is coincident with separability.
The resulting promotion of the \emph{negativity sphere} to an \emph{entanglement sphere} emphasizes that negativity serves, in this case, as a clear quantifier of experimentally relevant quantum correlations.
Spacelike separated detectors interacting locally with the  vacuum at regions carrying zero/non-zero negativity
will never/always be able to extract, in principle, entanglement from the vacuum.

%%%%%%%%%%%%%%%%%%%%%%%%%%%%%%%%%%%%%%%%%%%%%
\section{Numerical Calculation of \texorpdfstring{$\beta_{3D}$}{beta3D} and \texorpdfstring{$\gamma_{3D}$}{gamma3D}}
\label{app:3d}

In this appendix we elaborate on our calculation of the negativity between two pixelated spherical regions of  massless noninteracting lattice scalar field theory in 3D,
in particular the values of $\beta_{3D}=7.6(1)$ and $\gamma_{3D}=0.43(2)$, which are first calculations of these quantities.
In general, the calculation follows the same procedure that led to the determinations of $\beta_{1D, 2D}$ and $\gamma_{1D, 2D}$
presented in previous works, but there are  slight differences.
The geometry of the calculation is that of two equal sized, approximately spherical pixelated
regions of the lattice defining the two regions, $A$ and $B$,
that are separated along one of the Cartesian directions.
Each region has a diameter along an axis defined by $d_{\rm lat}$ lattice sites and of length $(d_{\rm lat}-1) a$, and are separated by a surface-to-surface distance of $(\tilde{r}+1) a$ and a distance between centers of $r a = (\tilde r + d_{\rm lat}-1) a$ lattice sites.
($d_{\rm lat}$ even or odd are both legitimate choices at finite lattice spacing that will produce the same continuum limit.)

Transverse dimensions are distinguished from the displacement direction.
While calculations of negativity can proceed by the computation of two-point correlation functions
followed by direct matrix diagonalization,
the dimensionality of the problem in higher dimensions and the condition number of the matrix
(from the polynomial scaling correlators giving rise to  an exponentially decreasing negativity),
motivate utilizing the discrete symmetries  of the system from the outset.
These include parity, reflections in the displacement plane and hypercubic symmetries.
Due to their observed dominant contribution to the negativity, we form the
$A_1^+$ irreps of the transverse H(D-1) group, defined by their radius from the transverse symmetry axis (including multiplicities where appropriate) for each distance along the displacement axis.
These coordinates are used to construct the $\hat G$ and $\hat H$ matrices.
For a given configuration, the calculation proceeds as follows:
\begin{enumerate}
\item
define the lattice sites in regions $A$ and $B$,
\item
determine the  displacement vectors for each pair of $A_1^+$ in the system for a given longitudinal displacement,
\item
numerically evaluate the integrals $G({\bf n})$ and $H^\Gamma({\bf n})$
(in the text) or retrieve the stored values, and form the appropriate symmetry combination,
\item
multiply  $\hat G$ and $\hat H^\Gamma$, and find the negative eigenvalues.
\end{enumerate}
\begin{figure}
\centering
\begin{minipage}{0.7\textwidth}
\includegraphics[width = 0.72\textwidth]{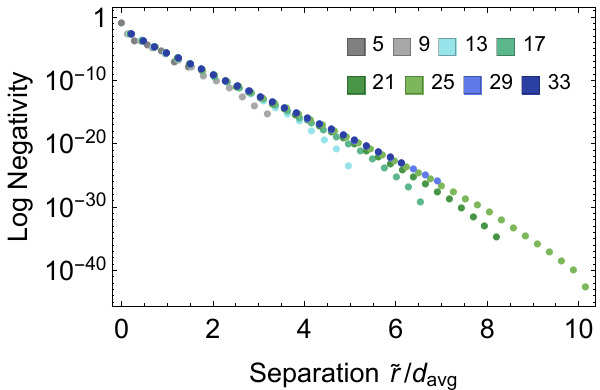}
\end{minipage}
\begin{minipage}{.20\textwidth}
  \centering
$\begin{array}{c|c}
\hline
\hline
1/d_{\rm lat} & \beta_{3D}(d) \\
\hline
\hline
 0.030 & 7.6901(89)   \\
 0.034  & 7.735(15)  \\
 0.040 & 7.732(15)    \\
 \rowcolor{Gray}
 0.048 & 8.014(65)  \\
 \rowcolor{Gray}
 0.059 & 8.38(11)   \\
 \rowcolor{Gray}
 0.077 & 9.70(16)  \\
 \rowcolor{Gray}
 0.111  & 10.01(19)   \\
 \hline
\hline
\end{array} $
 \end{minipage}
  \caption{
    (Left) The calculated log negativity for different pixelations of 3D spheres.  The values are provided in Table~\ref{tab:3Dnega}.
    (Right) The values of $\beta_{3D}^{({\rm eff})}$ determined for each pixelation, with the extrapolation to $1/d\rightarrow 0$ defining the continuum limit.
  }
  \label{fig:Negadata}
\end{figure}
The calculations that we present here were performed using a combination of {\tt Mathematica} and {\tt Matlab}, due to their different optimizations for the combinations of analytic and numeric functions we have employed.
In order to move beyond the calculations we show, more robust and scalable codes optimized for large-scale parallel computing
will be required.
The results of our calculations in 3D are provided in Table~\ref{tab:3Dnega} and displayed in Fig.~\ref{fig:Negadata}.

\begin{figure}
\footnotesize
\begin{minipage}{0.40\textwidth}
  \includegraphics[width = 0.65\textwidth]{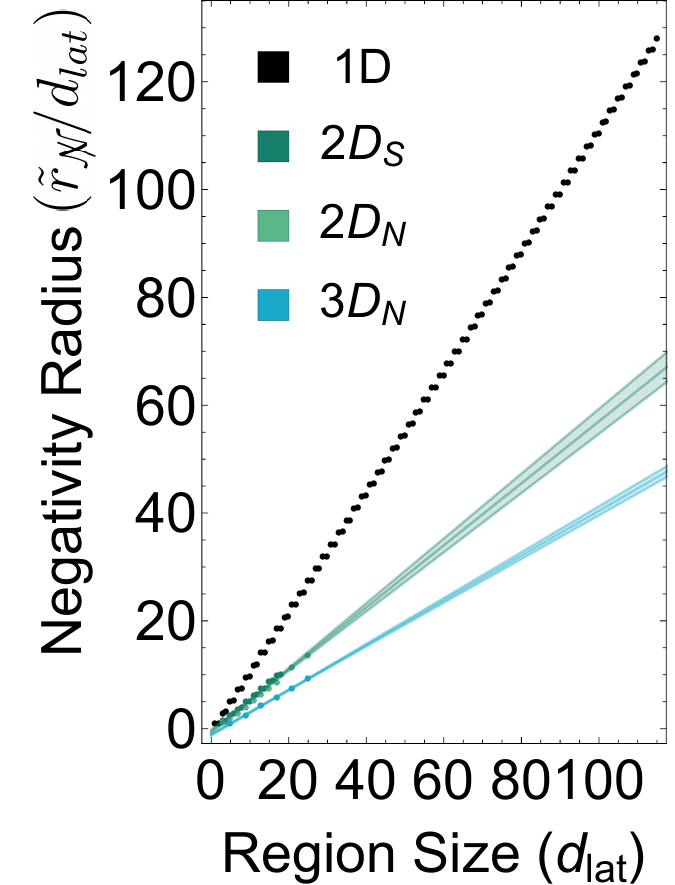}
\end{minipage}
\begin{minipage}{.20\textwidth}
  \centering
$\begin{array}{c|cc}
\hline
\hline
d_{\rm lat} & \tilde r_\Nslash/d_{\rm avg} & \tilde{r}_{\Nslash}/d_{\rm lat} \\
\hline
\hline
 5 & 1.46 & 1.0 \\
 9 & 3.20^{(+0.14)} & 2.56^{(+0.11)}\\
 13 & 4.97^{(+0.18)} & 4.31^{(+0.15)}\\
 17 & 6.55^{(+0.2)} & 5.83^{(+0.18)}\\
 21 & 8.21^{(+0.21)} & 7.57^{(+0.19)}\\
 25 & 10.16^{(+0.22)} & 9.32^{(+0.2)}\\
\end{array} $
 \end{minipage}
 \caption{
 (Left)
 The log negativity radius as a function of pixelation in 1D, 2D and 3D.
 The slopes provide $\gamma_{1D, 2D, 3D}$.
 The results shown for  1D and 2D are those from Ref.~\cite{Klco:2020rga}.
 (Right)
 Numerical locations of the last calculated non-zero negativity at the
 entanglement sphere.
 The uncertainties express the inexact determination of $\tilde{r}_{\Nslash}$ with the chosen separation sample density.
 }
 \label{fig:3Dextrapgamma}
\end{figure}

Leveraging techniques widely-used  in the analysis of lattice QCD calculations,
an effective mass is generated for
each point in each of the sets of results given in Table~\ref{tab:3Dnega}.
The effective masses tend to a constant from both small and large displacements,
with a range that increases  and oscillatory contributions that diminish with  denser pixelation of the spheres.
These effective masses  lend themselves to Fourier transformation to extract the asymptotic constant
with an associated uncertainty (the argument of the purely exponential behavior of the negativity)
for each level of pixelation.
The results of this extraction are presented in the table at the right of Fig.~\ref{fig:Negadata}.
For coarse pixelation, the range of non-vanishing negativity that we have computed is sufficiently limited that a robust estimate of the systematic errors are not practical, and we have been able to compute only three reliable values
of the argument of the exponential that can be used to extrapolate to the continuum with confidence.
A linear extrapolation of the three results in Fig.~\ref{fig:Negadata} with the smallest values of $1/d$ yields
$\beta_{3D}=7.6(1)$.
A higher precision result will require additional calculations at finer pixelation, which will enable extrapolations
using higher-order polynomials.
The present calculations, unlike our results in 2D, are the minimum that can be performed to obtain an extrapolated value.

The values of the points of vanishing negativity as a function of pixelation  determine the coefficient
$\gamma_{3D}$.
The results shown in Table~\ref{tab:3Dnega}
of the negativity as a function of separation
furnish the results given in the table at the right of Fig.~\ref{fig:3Dextrapgamma} for the (dimensionless) point of vanishing negativity.
Extracting the slope of these results with respect to $d_{\rm lat}$ yields a value of
$\gamma_{3D}=0.43(2)$.
It is interesting that the suppression of fluctuations in higher dimension has allowed the precision attained in 3D to be commensurate with that attained in 2D.
These numerical calculations provide early insight into the dependence of the vacuum entanglement structure on the spatial dimension of the field, yielding results sufficiently intriguing to motivate future calculations with higher precision and in higher dimension.

\begin{table}[!ht]
\footnotesize
\begin{minipage}[b]{.20\textwidth}
  \centering
\resizebox{2.5cm}{!}{
$\begin{array}{c|c}
\hline
\hline
\tilde{r} /d_{\rm avg} & {\cal N} \\
\hline
\hline
\multicolumn{2}{c}{d_{\rm lat} = 5 } \\
\hline
 0 & 1.19\times 10^{-1} \\
 2.93\times 10^{-1} & 2.00\times 10^{-4} \\
 5.86\times 10^{-1} & 3.97\times 10^{-5} \\
 8.79\times 10^{-1} & 4.31\times 10^{-6} \\
 1.17 & 1.03\times 10^{-7} \\
 1.46 & 1.64\times 10^{-8} \\
\hline
\multicolumn{2}{c}{d_{\rm lat} = 9 } \\
\hline
 1.39\times 10^{-1} & 2.88\times 10^{-3} \\
 4.17\times 10^{-1} & 1.90\times 10^{-4} \\
 6.96\times 10^{-1} & 1.32\times 10^{-5} \\
 9.74\times 10^{-1} & 1.70\times 10^{-6} \\
 1.25 & 2.19\times 10^{-7} \\
 1.53 & 1.37\times 10^{-8} \\
 1.81 & 6.45\times 10^{-10} \\
 2.09 & 8.72\times 10^{-11} \\
 2.37 & 7.48\times 10^{-12} \\
 2.64 & 2.51\times 10^{-13} \\
 2.92 & 9.87\times 10^{-15} \\
 3.2 & 6.30\times 10^{-16} \\
\hline
\multicolumn{2}{c}{d_{\rm lat} = 13 } \\
\hline
 1.77\times 10^{-1} & 2.44\times 10^{-3} \\
 4.44\times 10^{-1} & 2.43\times 10^{-4} \\
 7.1\times 10^{-1} & 1.93\times 10^{-5} \\
 9.76\times 10^{-1} & 2.91\times 10^{-6} \\
 1.24 & 2.94\times 10^{-7} \\
 1.51 & 3.84\times 10^{-8} \\
 1.77 & 5.32\times 10^{-9} \\
 2.04 & 5.33\times 10^{-10} \\
 2.31 & 5.64\times 10^{-11} \\
 2.57 & 6.60\times 10^{-12} \\
 2.84 & 7.06\times 10^{-13} \\
 3.11 & 6.16\times 10^{-14} \\
 3.37 & 4.67\times 10^{-15} \\
 3.64 & 5.82\times 10^{-16} \\
 3.9 & 5.43\times 10^{-17} \\
 4.17 & 1.05\times 10^{-18} \\
 4.44 & 4.08\times 10^{-20} \\
 4.7 & 1.90\times 10^{-21} \\
 4.97 & 3.55\times 10^{-24} \\
 \end{array} $
 }
 \end{minipage}\qquad
\begin{minipage}[b]{.20\textwidth}
  \centering
\resizebox{2.5cm}{!}{
$\begin{array}{c|c}
\hline
\hline
\tilde{r} /d_{\rm avg} & {\cal N} \\
\hline
\hline
\multicolumn{2}{c}{d_{\rm lat} = 17 }\\
\hline
 1.99\times 10^{-1} & 2.32\times 10^{-3} \\
 4.63\times 10^{-1} & 1.83\times 10^{-4} \\
 7.28\times 10^{-1} & 1.97\times 10^{-5} \\
 9.93\times 10^{-1} & 2.34\times 10^{-6} \\
 1.26 & 2.68\times 10^{-7} \\
 1.52 & 3.97\times 10^{-8} \\
 1.79 & 4.18\times 10^{-9} \\
 2.05 & 4.68\times 10^{-10} \\
 2.32 & 7.23\times 10^{-11} \\
 2.58 & 8.65\times 10^{-12} \\
 2.85 & 1.01\times 10^{-12} \\
 3.11 & 1.28\times 10^{-13} \\
 3.38 & 1.52\times 10^{-14} \\
 3.64 & 1.43\times 10^{-15} \\
 3.91 & 1.58\times 10^{-16} \\
 4.17 & 2.09\times 10^{-17} \\
 4.44 & 1.65\times 10^{-18} \\
 4.7 & 1.55\times 10^{-19} \\
 4.97 & 9.53\times 10^{-21} \\
 5.23 & 7.83\times 10^{-22} \\
 5.5 & 4.67\times 10^{-23} \\
 5.76 & 1.60\times 10^{-24} \\
 6.03 & 7.07\times 10^{-26} \\
 6.29 & 1.84\times 10^{-27} \\
 6.55 & 6.29\times 10^{-30} \\
\hline
\multicolumn{2}{c}{d_{\rm lat} = 21 } \\
\hline
 2.07\times 10^{-1} & 2.64\times 10^{-3} \\
 4.65\times 10^{-1} & 1.84\times 10^{-4} \\
 7.23\times 10^{-1} & 2.24\times 10^{-5} \\
 9.81\times 10^{-1} & 2.45\times 10^{-6} \\
 1.24 & 3.30\times 10^{-7} \\
 1.5 & 4.42\times 10^{-8} \\
 1.76 & 4.79\times 10^{-9} \\
 2.01 & 7.28\times 10^{-10} \\
 2.27 & 9.43\times 10^{-11} \\
 2.53 & 1.16\times 10^{-11} \\
 2.79 & 1.83\times 10^{-12} \\
 3.05 & 2.27\times 10^{-13} \\
 3.31 & 2.49\times 10^{-14} \\
 3.56 & 3.53\times 10^{-15} \\
 3.82 & 4.55\times 10^{-16} \\
 4.08 & 5.46\times 10^{-17} \\
 4.34 & 6.83\times 10^{-18} \\
 4.6 & 8.56\times 10^{-19} \\
 4.86 & 9.23\times 10^{-20} \\
 5.11 & 9.97\times 10^{-21} \\
 5.37 & 9.97\times 10^{-22} \\
 5.63 & 9.76\times 10^{-23} \\
 5.89 & 8.84\times 10^{-24} \\
 6.15 & 9.80\times 10^{-25} \\
 6.41 & 7.34\times 10^{-26} \\
 6.66 & 4.72\times 10^{-27} \\
 6.92 & 2.69\times 10^{-28} \\
 7.18 & 2.05\times 10^{-29} \\
 7.44 & 1.02\times 10^{-30} \\
 7.7 & 3.04\times 10^{-32} \\
 7.96 & 1.37\times 10^{-33} \\
 8.21 & 2.41\times 10^{-35} \\
 \end{array} $
 }
 \end{minipage}\qquad
\begin{minipage}[b]{.20\textwidth}
  \centering
\resizebox{2.5cm}{!}{
$\begin{array}{c|c}
\hline
\hline
\tilde{r} /d_{\rm avg} & {\cal N} \\
\hline
\hline
\multicolumn{2}{c}{d_{\rm lat} = 25 }\\
\hline
 2.18\times 10^{-1} & 2.63\times 10^{-3} \\
 4.79\times 10^{-1} & 1.76\times 10^{-4} \\
 7.41\times 10^{-1} & 2.22\times 10^{-5} \\
 1. & 2.34\times 10^{-6} \\
 1.26 & 3.34\times 10^{-7} \\
 1.53 & 3.89\times 10^{-8} \\
 1.79 & 5.17\times 10^{-9} \\
 2.05 & 7.45\times 10^{-10} \\
 2.31 & 8.57\times 10^{-11} \\
 2.57 & 1.31\times 10^{-11} \\
 2.83 & 1.83\times 10^{-12} \\
 3.09 & 2.17\times 10^{-13} \\
 3.36 & 3.20\times 10^{-14} \\
 3.62 & 4.42\times 10^{-15} \\
 3.88 & 5.13\times 10^{-16} \\
 4.14 & 6.94\times 10^{-17} \\
 4.4 & 9.14\times 10^{-18} \\
 4.66 & 1.22\times 10^{-18} \\
 4.93 & 1.47\times 10^{-19} \\
 5.19 & 1.82\times 10^{-20} \\
 5.45 & 2.03\times 10^{-21} \\
 5.71 & 2.23\times 10^{-22} \\
 5.97 & 2.60\times 10^{-23} \\
 6.23 & 2.66\times 10^{-24} \\
 6.49 & 2.93\times 10^{-25} \\
 6.76 & 3.07\times 10^{-26} \\
 7.02 & 2.89\times 10^{-27} \\
 7.28 & 3.09\times 10^{-28} \\
 7.54 & 2.26\times 10^{-29} \\
 7.8 & 2.71\times 10^{-30} \\
 8.06 & 2.19\times 10^{-31} \\
 8.32 & 1.17\times 10^{-32} \\
 8.59 & 6.11\times 10^{-34} \\
 8.85 & 3.84\times 10^{-35} \\
 9.11 & 2.11\times 10^{-36} \\
 9.37 & 9.88\times 10^{-38} \\
 9.63 & 4.29\times 10^{-39} \\
 9.89 & 1.54\times 10^{-40} \\
 1.02\times 10^1 & 3.35\times 10^{-43} \\
\end{array} $
 }
 \end{minipage}\qquad
\begin{minipage}[b]{.20\textwidth}
  \centering
\resizebox{2.5cm}{!}{
$\begin{array}{c|c}
\hline
\hline
\tilde{r} /d_{\rm avg} & {\cal N} \\
\hline
\hline
\multicolumn{2}{c}{d_{\rm lat} = 29 } \\
\hline
 2.21\times 10^{-1} & 2.80\times 10^{-3} \\
 4.78\times 10^{-1} & 1.92\times 10^{-4} \\
 7.36\times 10^{-1} & 2.34\times 10^{-5} \\
 9.93\times 10^{-1} & 2.60\times 10^{-6} \\
 1.25 & 3.67\times 10^{-7} \\
 1.51 & 4.23\times 10^{-8} \\
 1.77 & 6.29\times 10^{-9} \\
 2.02 & 8.45\times 10^{-10} \\
 2.28 & 1.08\times 10^{-10} \\
 2.54 & 1.67\times 10^{-11} \\
 2.8 & 2.20\times 10^{-12} \\
 3.05 & 3.02\times 10^{-13} \\
 3.31 & 4.44\times 10^{-14} \\
 3.57 & 5.79\times 10^{-15} \\
 3.83 & 7.79\times 10^{-16} \\
 4.08 & 1.10\times 10^{-16} \\
 4.34 & 1.42\times 10^{-17} \\
 4.6 & 1.93\times 10^{-18} \\
 4.86 & 2.62\times 10^{-19} \\
 5.11 & 3.32\times 10^{-20} \\
 5.37 & 4.36\times 10^{-21} \\
 5.63 & 5.82\times 10^{-22} \\
 5.89 & 7.26\times 10^{-23} \\
 6.14 & 9.13\times 10^{-24} \\
 6.4 & 1.20\times 10^{-24} \\
 6.66 & 1.50\times 10^{-25} \\
 6.92 & 1.60\times 10^{-26} \\
 \hline
\multicolumn{2}{c}{d_{\rm lat} = 33 } \\
\hline
 2.25\times 10^{-1} & 2.71\times 10^{-3} \\
 4.82\times 10^{-1} & 1.90\times 10^{-4} \\
 7.39\times 10^{-1} & 2.24\times 10^{-5} \\
 9.97\times 10^{-1} & 2.58\times 10^{-6} \\
 1.25 & 3.47\times 10^{-7} \\
 1.51 & 4.20\times 10^{-8} \\
 1.77 & 6.28\times 10^{-9} \\
 2.03 & 7.75\times 10^{-10} \\
 2.28 & 1.10\times 10^{-10} \\
 2.54 & 1.61\times 10^{-11} \\
 2.8 & 2.08\times 10^{-12} \\
 3.05 & 3.17\times 10^{-13} \\
 3.31 & 4.37\times 10^{-14} \\
 3.57 & 5.71\times 10^{-15} \\
 3.83 & 8.27\times 10^{-16} \\
 4.08 & 1.13\times 10^{-16} \\
 4.34 & 1.53\times 10^{-17} \\
 4.6 & 2.15\times 10^{-18} \\
 4.85 & 2.87\times 10^{-19} \\
 5.11 & 3.92\times 10^{-20} \\
 5.37 & 5.28\times 10^{-21} \\
 5.63 & 6.77\times 10^{-22} \\
 5.88 & 8.52\times 10^{-23} \\
 6.14 & 1.13\times 10^{-23} \\
\end{array} $ }
 \end{minipage}
 \caption{
 The numerically determined negativity between 3D pixelated spherical lattice regions as a function of normalized separation.
 The columns of finer-pixelation  (d = 29, 33) do not show results throughout the entire range of non-vanishing negativity.
 }
 \label{tab:3Dnega}
\end{table}
%

%%%%%%%%%%%%%%%%%%%%%%%%%%%%%%%
\section{Dispersion Relation Improvement in One Dimension}
\label{app:dispersion}
The numerical lattice scalar field theory
calculations performed in this work, and the results of previously published works (e.g., Refs.~\cite{Marcovitch:2008sxc,Klco:2020rga}), have employed the
simplest discretization of the lattice energy-momentum relation in which the continuum momentum operator is approximated
by a nearest neighbor finite-difference operation on lattice sites.
It is helpful to quantify the sensitivity of the physics results we have presented in this paper to the choice of lattice discretization
of this gradient operator.
It is well known how to systematically improve the dispersion relation by introducing higher order operators, suppressed by positive powers of the lattice spacing,
to remove higher-order contributions in momentum and to converge toward the desired $E^2=k^2 + m^2$ of special relativity.
However, for a finite number of lattice sites, only a finite number of independent finite-difference operators may be constructed, and each order of improvement corresponds to further smearing of the gradient operator over an increasing region of the lattice.
To explore the impact of the discretization, we compare results in one dimension obtained with lattice dispersion relations of the form:
\begin{eqnarray}
&& m^2  + 4\sin^2 \left(k/2\right)\ \rightarrow\ m^2 + k^2 -\frac{k^4}{ 12} + \frac{k^6}{ 360} - \frac{k^8}{ 20160} + {\cal O}\left(k^{10}\right)
\nonumber\\
&& m^2  + 4\sin^2 \left(k/2\right)+ \frac{4}{ 3}\sin^4 \left(k/2\right)\ \rightarrow\ m^2 + k^2  - \frac{k^6}{ 90} + \frac{k^8}{ 1008} + {\cal O}\left(k^{10}\right)
\ \ \ ,
\label{app:improvedD}
\end{eqnarray}
which have the same low-momentum behavior up to polynomial corrections and differ by $\mathcal{O}(1)$ at the edge of the Brillouin zone.
These correspond to different discretizations while preserving the infrared behavior of the lattice scalar field theory,
and correspond to leading order and  ${\cal O}(k^{4})$-improved lattice
actions for the non-interacting lattice theory.
While it might be tempting to consider the full dispersion relation directly,  with the finite number of lattice sites in the system such an operator cannot be accurately obtained from a finite number of finite-difference operators.
Different discretizations of the scalar field that preserve the energy-momentum relation correspond to including smearings of the gradient operator, and can be included as higher-order terms in the dispersion relation.
As the goal of this Appendix is to illustrate the impact of modified UV physics on the entanglement structure of the field and because modified lattice actions are closely related to smearing for bosonic fields, the first order improvement of the dispersion relation is chosen to favor the relative impact of the UV modification.

In this Appendix, the negativity between two vacuum regions is calculated through two-point functions in which the na\"ive lattice dispersion relations (see  Appendix~\ref{app:correlationfunctions}), are replaced by the $k^4$-improved relations in Eq.~(\ref{app:improvedD}).
Numerical evaluations of the correlation functions become increasingly weighted to low momentum at large distances, with the higher-order improvements making a decreasing contribution to each of the integrals.
On the other hand, the long-distance negativity becomes increasingly sensitive to the short-distance behavior of the two-point functions and does explicitly probe these higher-order improvements, consistent with the main thesis of this article.
Together, these observations indicate the presence of large cancelations between polynomially- and logarithmically-varying correlation functions at long distances to produce an exponentially decaying residual sensitive to the short-distance physics.
Table~\ref{tab:dispresults} shows the results obtained
for the radius of the negativity sphere for different region sizes
using the improved lattice dispersion relations.
\begin{figure}
  \includegraphics[width=0.8\columnwidth]{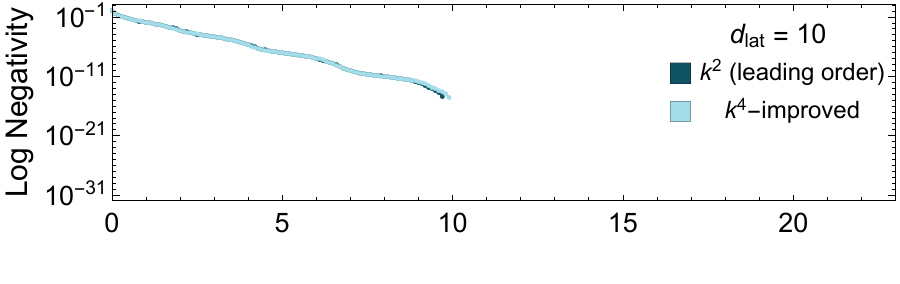}
  \includegraphics[width=0.8\columnwidth]{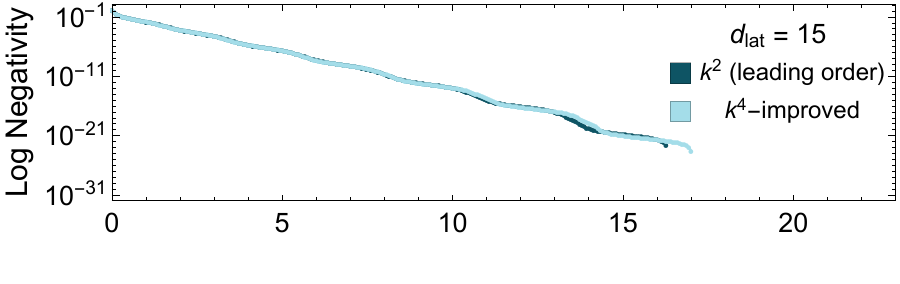}
  \includegraphics[width=0.8\columnwidth]{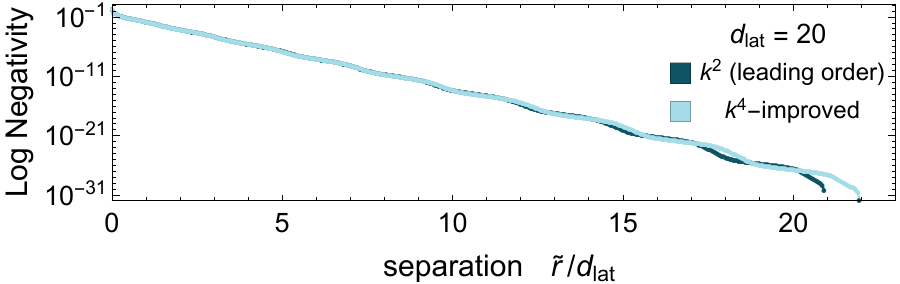}
  \caption{The logarithmic negativity calculated with two different dispersion relations ($k^2$ leading order and $k^4$-improved) for one-dimensional region lengths of 10(top), 15(middle), and 20(bottom) lattice sites.
  Sensitivity to the UV modification appears at large spatial separation, as discussed in the main text.
   }
  \label{fig:negadisp}
\end{figure}
The negativity between regions of different sizes are shown in Fig.~\ref{fig:negadisp} for two different dispersion relations.
Away from the continuum limit, the negativity exhibits oscillations that depend upon the dispersion relation,
and as such lead to fluctuations in the separability radius but do not change the scale setting the exponential dependence on separation.
As discussed, the negativity at smaller separations is less sensitive to the high-momentum behavior, consistent with a smaller dependence on the dispersion relation improvement.

\begin{table}
  \begin{tabular}{c|cc}
  \multicolumn{3}{c}{$k^2$} \\
  \hline
  \hline
  $d_{lat}$ & $ \tilde{r}_{\Nslash} $ & $\tilde{r}_{\Nslash}/d_{lat}$ \\
  \hline
  \hline
1 & 1 & 1.00 \\
2 & 2 & 1.00 \\
3 & 9 & 3.00 \\
4 & 13 & 3.25 \\
5 & 26 & 5.20 \\
6 & 32 & 5.33 \\
7 & 52 & 7.43 \\
8 & 61 & 7.63 \\
9 & 87 & 9.67 \\
10 & 98 & 9.80 \\
\end{tabular}
\begin{tabular}{c|cc}
\multicolumn{3}{c}{} \\
\multicolumn{3}{c}{} \\
\multicolumn{3}{c}{} \\
11 & 131 & 11.90 \\
12 & 144 & 12.00 \\
13 & 184 & 14.20 \\
14 & 199 & 14.20 \\
15 & 245 & 16.30 \\
16 & 264 & 16.50 \\
17 & 316 & 18.60 \\
18 & 337 & 18.70 \\
19 & 395 & 20.80 \\
20 & 419 & 21.00 \\
\hline
\hline
  \end{tabular}
  \qquad
    \begin{tabular}{c|cc}
  \multicolumn{3}{c}{$k^4$} \\
  \hline
  \hline
  $d_{lat}$ & $ \tilde{r}_{\Nslash} $ & $\tilde{r}_{\Nslash}/d_{lat}$ \\
  \hline
  \hline
1 & 1 & 1.00 \\
2 & 2 & 1.00 \\
3 & 9 & 3.00 \\
4 & 12 & 3.00 \\
5 & 26 & 5.20 \\
6 & 32 & 5.33 \\
7 & 53 & 7.57 \\
8 & 61 & 7.63 \\
9 & 89 & 9.89 \\
10 & 100 & 10.00 \\
\end{tabular}
\begin{tabular}{c|cc}
\multicolumn{3}{c}{} \\
\multicolumn{3}{c}{} \\
\multicolumn{3}{c}{} \\
11 & 135 & 12.30 \\
12 & 148 & 12.30 \\
13 & 190 & 14.60 \\
14 & 207 & 14.80 \\
15 & 256 & 17.10 \\
16 & 275 & 17.20 \\
17 & 331 & 19.50 \\
18 & 352 & 19.60 \\
19 & 415 & 21.80 \\
20 & 440 & 22.00 \\
  \hline
  \hline
  \end{tabular}
  \caption{Negativity sphere radii for regions of the one-dimensional massless scalar field with a nearest neighbor lattice action ($k^2$) and order $k^4$-improvement of the dispersion relation. }
  \label{tab:dispresults}
\end{table}

\end{document}